\documentclass[12pt]{article}

\usepackage[dvips]{color}                 
\usepackage{graphicx}                     
\usepackage{amssymb}
\usepackage{amsmath}

\usepackage{xspace}                       

\renewcommand{\today}{\ifcase\day\or 1st\or 2nd\or 3rd\or 4th\or 5th\or 6th\or
  7th\or 8th\or 9th\or 10th\or 11th\or 12th\or 13th\or 14th\or 15th\or 16th\or
  17th\or 18th\or 19th\or 20th\or 21st\or 22nd\or 23rd\or 24th\or 25th\or
  26th\or 27th\or 28th\or 29th\or 30th\or 31st\fi~\ifcase\month\or January\or
  February\or March\or April\or May\or June\or July\or August\or September\or
  October\or November\or December\fi \space \number\year}

\newcommand{\journal}[4]{{#1}\textbf{#2},
  #4 (#3)}

\newcommand{\EPJA}{\textit{Eur.\  Phys.\ J.\ }\textbf{A}}

\newcommand{\NPA}{\textit{Nucl.\ Phys.\ }\textbf{A}}
\newcommand{\NPB}{\textit{Nucl.\ Phys.\ }\textbf{B}}
\newcommand{\PLB}{\textit{Phys.\ Lett.\ }\textbf{B}}
\newcommand{\PR}{\textit{Phys.\ Rev.\ }}

\newcommand{\PRC}{\PR\textbf{C}}
\newcommand{\PRD}{\PR\textbf{D}}
\newcommand{\PRL}{\PR\textit{Lett.\ }}

\newcommand{\beq}{\begin{equation}}
\newcommand{\eeq}{\end{equation}}

\newcommand{\mytitle}[1]{\begin{center}\LARGE{\textbf{#1}}\end{center}}
\newcommand{\myauthor}[1]{\textbf{#1}}
\newcommand{\myaddress}[1]{\textit{#1}}
\newcommand{\mypreprint}[1]{\begin{flushright}#1\end{flushright}}

\begin{document}
%

\begin{titlepage}
  
  \mypreprint{\hfill    nucl-th/0405077\\
    TUM-T39-04-06}

  \vspace*{0.1cm}
  
  \mytitle{Explicit Delta(1232) Degrees of Freedom in Compton Scattering off 
           the Deuteron}

  \vspace*{0.3cm}

\begin{center}
  \myauthor{Robert P. Hildebrandt$^{a,}$}\footnote{
    Email: rhildebr@physik.tu-muenchen.de}, 
  \myauthor{Harald W.\ Grie\3hammer$^{a,b,}$}\footnote{
    Email: hgrie@physik.tu-muenchen.de; permanent address: a}, \\
  \myauthor{Thomas R.~Hemmert$^{a,b,}$}\footnote{
    Email: themmert@physik.tu-muenchen.de; permanent address: a} and 
  \myauthor{Daniel R.~Phillips$^{c,}$}\footnote{
    Email: phillips@phy.ohiou.edu}\\[2ex]

  \vspace*{0.5cm}
  
  \myaddress{$^a$
    Institut f{\"u}r Theoretische Physik (T39), Physik-Department,\\
    Technische Universit{\"a}t M{\"u}nchen, D-85747 Garching, Germany}
  \\[2ex]
  \myaddress{$^b$ ECT*, Villa Tambosi, I-38050 Villazzano (Trento), Italy}
  \\[2ex]
  \myaddress{$^c$ Department of Physics and Astronomy, Ohio University, 
    Athens, OH 45701}

\end{center}

\vspace*{0.5cm}

\begin{abstract}
We examine elastic Compton scattering off the deuteron for photon energies 
between 50~MeV and 100~MeV in
the framework of chiral effective field theories to next-to-leading order. 
We compare one theoretical scheme with only pions and nucleons 
as explicit degrees of freedom to another in which the $\Delta(1232)$ resonance
is treated as an explicit degree of freedom.
Whereas pion degrees of freedom suffice to describe the 
experimental data measured at about 70 MeV, the explicit $\Delta(1232)$  
gives important contributions that help to reproduce the angular dependence 
at higher energies. 
The static isoscalar dipole polarizabilities $\alpha_E^s$ and 
$\beta_M^s$ are fitted 
to the available data, giving results for the neutron polarizabilities  
$\alpha_E^n=(14.2\pm2.0\,(\mathrm{stat})\pm1.9\,(\mathrm{syst}))
            \cdot 10^{-4}\;\mathrm{fm}^3$,
$\beta_M^n =( 1.8\pm2.2\,(\mathrm{stat})\pm0.3\,(\mathrm{syst}))
            \cdot 10^{-4}\;\mathrm{fm}^3$.
These values are in good agreement with previous experimental analyses. 
Comparing them to the well-known proton values we 
conclude that there is currently no evidence for significant differences 
between the proton and neutron electromagnetic dipole polarizabilities.
\end{abstract}
\vskip 1.0cm
\noindent
\begin{tabular}{rl}
Suggested PACS numbers:& 13.40.-f, 13.60.Fz, 25.20.-x, 21.45.+v
\\[1ex]
Suggested Keywords: &\begin{minipage}[t]{11cm}
                    Effective Field Theory, Compton Scattering,\\
                    Deuteron, Delta Resonance, Nucleon Polarizabilities. 
                    \end{minipage}
\end{tabular}

\vskip 1.0cm

\end{titlepage}

\setcounter{page}{2} \setcounter{footnote}{0} \newpage

%

\section{Introduction}
\setcounter{equation}{0}
\label{sec:introduction}
The structure of protons and neutrons as analyzed with electromagnetic probes 
has been under experimental and theoretical investigation for a number of 
decades.
In Compton scattering the external electromagnetic field of the photon
attempts to deform the nucleon. 
The electromagnetic polarizabilities provide a measure of the global 
resistance of the
nucleon's internal degrees of freedom against displacement in an external 
electric or magnetic field, which makes 
them an excellent tool to study the sub-nucleonic degrees of freedom. 
If one defines the polarizabilities via a multipole expansion in Compton 
scattering, these quantities are energy dependent.
The physics connected with this energy dependence is discussed in 
Refs.~\cite{GH01,HGHP}. 
In this work we are mainly concerned with the static values, which we 
therefore denote as \textit{the} polarizabilities for simplicity.  
Experimentally the best-known nucleon polarizabilities are the static 
electric and magnetic dipole polarizabilities of the proton, $\alpha_E^p$ and 
$\beta_M^p$. The results of the global fit to the wealth of Compton 
scattering data on the proton given in Ref.~\cite{Olmos} are 
\begin{align}
\alpha_E^p&=(12.1\pm0.3\,(\mathrm{stat})\mp0.4\,(\mathrm{syst})
            \pm0.3\,(\mathrm{model}))\cdot 10^{-4}\;\mathrm{fm}^3\,,\nonumber\\
\beta_M^p &=( 1.6\pm0.4\,(\mathrm{stat})\pm0.4\,(\mathrm{syst})
            \pm0.4\,(\mathrm{model}))\cdot 10^{-4}\;\mathrm{fm}^3\,.
\label{expp}
\end{align}
The values from a Baldin Sum Rule constrained fit of the proton Compton 
data within the framework that we use in this work, reported in
Ref.~\cite{HGHP}, agree within error bars with~(\ref{expp}):
\begin{equation}
\alpha_E^p=(11.04\pm1.36)\cdot 10^{-4}\;\mathrm{fm}^3\,,\;\;\;\;
\beta_M^p =( 2.76\mp1.36)\cdot 10^{-4}\;\mathrm{fm}^3\,.
\label{exppHGHP}
\end{equation}
The errors displayed in Eq.~(\ref{exppHGHP}) are only statistical. For the 
fit, the central value of the Baldin Sum Rule 
$\alpha_E^p+\beta_M^p=(13.8\pm0.4)\cdot 10^{-4}\;\mathrm{fm}^3$~\cite{Olmos} 
has been used.

Due to the lack of stable single-neutron targets for Compton scattering it is 
much harder to access the neutron polarizabilities experimentally. 
An experiment on quasi-free Compton scattering from the proton 
and neutron bound in the deuteron~\cite{Kossert} gives results for the neutron 
polarizabilities which suggest 
very small isovectorial components\footnote{The isovector polarizabilities are 
defined as $\alpha_E^v\equiv\frac{1}{2}\,(\alpha_E^p-\alpha_E^n)$, 
           $\beta_ M^v\equiv\frac{1}{2}\,(\beta _M^p-\beta _M^n)$.} when 
compared to Eqs.~(\ref{expp},\ref{exppHGHP}):
\begin{align}
\alpha_E^n&=(12.5\pm1.8\,(\mathrm{stat})\,^{+1.1}_{-0.6}\,(\mathrm{syst})
            \pm1.1\,(\mathrm{model}))\cdot 10^{-4}\;\mathrm{fm}^3\nonumber\\
\beta_M^n &=( 2.7\mp1.8\,(\mathrm{stat})\,^{+0.6}_{-1.1}\,(\mathrm{syst})
            \mp1.1\,(\mathrm{model}))\cdot 10^{-4}\;\mathrm{fm}^3
\label{expn}
\end{align}
A similar observation for $\alpha_E^n$ has been made in~\cite{Schmied}, where 
the scattering of neutrons on lead was measured:
\begin{equation}
\alpha_E^n=(12.6\pm2.5)\cdot 10^{-4}\;\mathrm{fm}^3
\label{expSchmied}
\end{equation}
However, the precision of this result has been questioned by the authors of 
\cite{Enik}. Their estimate of the correct range for the result 
from~\cite{Schmied} is $7\leq\alpha_E^n\leq19$.
On the other hand, another experiment~\cite{Koester}, using the same 
technique, gives a completely  different result:
\beq
\alpha_E^n=(0.6\pm5.0)\cdot 10^{-4}\;\mathrm{fm}^3
\eeq

On the theory side Chiral Perturbation Theory predicts that the proton and 
neutron 
polarizabilities are equal at leading-one-loop order \cite{BKKM}, 
since the pion loops 
that generate these contributions are isoscalar in nature. The absence of 
large isovector pieces in $\alpha_E$ and $\beta_M$ therefore is in accord 
with this picture.

Another possible way to determine the neutron polarizabilities is elastic 
low-energy Compton scattering from light nuclei, e.g. from the deuteron. 
Several experiments have already been performed~\cite{Lucas,Lund,Hornidge} 
and further proposals exist -- e.g. Compton scattering on the  
deuteron or $\mathrm{He}^3$ at TUNL/HI$\gamma$S \cite{Gao} and on 
deuteron targets at MAXlab~\cite{Schroeder}. 
From a theorist's point of view, extracting the neutron 
polarizabilities from elastic $\gamma d$ scattering 
requires an accurate description of the nucleon structure  and the 
dynamics of the low-energy degrees of freedom within the deuteron, as one has 
to correct for the proton polarizabilities and meson-exchange effects. A first 
attempt to fit the isoscalar polarizabilities 
$\alpha_E^s\equiv\frac{1}{2}\,(\alpha_E^p+\alpha_E^n)$, 
$\beta_ M^s\equiv\frac{1}{2}\,(\beta _M^p+\beta _M^n)$ to the elastic 
deuteron Compton scattering data from~\cite{Lucas, Hornidge} has been made 
in~\cite{Lvov}. The extracted neutron polarizabilities 
$\alpha_E^n=( 9.0\pm3.0)\cdot 10^{-4}\;\mathrm{fm}^3$,
$\beta_ M^n=(11.0\pm3.0)\cdot 10^{-4}\;\mathrm{fm}^3$
indicate the possibility of a rather \textit{large} isovector part. 
On the other hand comparison 
of the elastic deuteron Compton calculation of Ref.~\cite{Miller} with the 
data from~\cite{Lucas} is in good agreement with nearly vanishing isovector 
polarizabilities: 
$\alpha_E^n=(12.0\pm4.0)\cdot 10^{-4}\;\mathrm{fm}^3$,
$\beta_ M^n=( 2.0\pm4.0)\cdot 10^{-4}\;\mathrm{fm}^3$, albeit within rather 
large error bars. 

It is obvious from these partly 
contradictory results that there is still a lot of work to be 
done in order to have reliable values for 
$\alpha_E^n$ and $\beta_M^n$.
In general,
Chiral Effective Field Theory provides a consistent, controlled framework 
for elastic $\gamma d$ scattering within 
which nucleon effects can be disentangled from meson-exchange currents, 
deuteron binding, etc. It therefore gives valuable contributions to the 
ongoing discussion on the neutron polarizabilities. 
This work aims for an improved description of elastic deuteron Compton data 
at $\omega\sim50$-100~MeV, 
compared to the calculations presented in \cite{Lvov,Miller}, which cannot 
match the data in this regime.

Our work is based on the calculation of Refs.~\cite{Phillips,McGPhil}, 
where Compton 
scattering off the deuteron was examined for photon energies $\omega$ ranging 
from 50~MeV to 100~MeV. 
The central values for the isoscalar polarizabilities, derived in the 
recent analysis~\cite{McGPhil} of the data from~\cite{Lucas, Lund, Hornidge}
are
\begin{align}
\alpha_E^s&=(8.9\pm1.5)^{+4.5}_{-0.9}\cdot 10^{-4}\;\mathrm{fm}^3\;,\nonumber\\
\beta_ M^s&=(2.2\pm1.5)^{+1.2}_{-0.9}\cdot 10^{-4}\;\mathrm{fm}^3\;.
\end{align}
Comparing with Eq.~(\ref{expp}), these results indicate
a small isovector magnetic polarizability, but signal
the possibility of a rather large $\alpha_E^v$.
However, the range for  
$\alpha_E^s$ and $\beta_ M^s$ quoted in~\cite{McGPhil} is rather large:
$\alpha_E^s=(8.0-13.6)\cdot 10^{-4}\;\mathrm{fm}^3$, 
$\beta_ M^s=(1.3- 3.4)\cdot 10^{-4}\;\mathrm{fm}^3$.
The authors of Ref.~\cite{Phillips} 
followed Weinberg's proposal to calculate the irreducible kernel for the 
$\gamma N N\rightarrow \gamma N N$ process in Heavy Baryon Chiral 
Perturbation Theory (HB$\chi$PT) and then fold this 
with external deuteron wave functions such as Nijm93~\cite{Nijm}, 
CD-Bonn~\cite{Bonn} or AV18~\cite{AV18}. Proceeding in 
this fashion means working within an Effective Field Theory in which only 
nucleons and pions are active degrees of freedom. This ``hybrid'' approach 
has proven quite successful in describing $\pi d$~\cite{Beane}, 
$e^- d$~\cite{Danieled}, 
and even $\gamma d$~\cite{Phillips,McGPhil} scattering.
In this work we extend the 
calculation of Ref.~\cite{Phillips} to an Effective Field Theory which 
includes the $\Delta(1232)$ 
resonance of the nucleon as an additional explicit degree of freedom.
The advantage of our approach with respect 
to the NNLO calculation of Ref.~\cite{McGPhil} is that we identify the physics 
hidden in some of the short-distance parameters there. As we shall see, this 
is particularly important for quantities such as the magnetic 
polarizability $\beta_M$, where the $\Delta(1232)$ plays an important 
dynamical role.
The huge influence of the $\Delta(1232)$ in single-nucleon Compton 
scattering -- especially in the backward direction -- is a well-known fact 
(see e.g.~\cite{HGHP}).
We note that the importance of the $\Delta$-contributions is due to the strong 
paramagnetic $M1$ coupling of the photon to the 
$N\rightarrow\Delta$ transition, visible already far below resonance
(cf.~\cite{HGHP}). 
It is therefore interesting to also investigate the role of these degrees of 
freedom in elastic $\gamma d$ scattering, 
and this is the main focus of our work.

The idea of an extension of HB$\chi$PT that includes explicit $\Delta$ degrees 
of freedom has its origin in the early 1990's~\cite{Jenkins}. 
When including the $\Delta(1232)$ explicitly in $\chi$EFT
one needs to specify how the $\Delta N$ mass splitting $\Delta_0$ is 
treated in the power counting. Here we use the so-called Small Scale Expansion
(SSE)~\cite{HHKLett}. We note that there also exist alternative approaches 
for Chiral Effective Field Theories with explicit $\pi$, $N$ and $\Delta$ 
degrees of freedom, e.g. the $\delta$-expansion~\cite{deltaexp}, which was 
recently shown to describe $\gamma p$ cross section 
data well in an energy range from $\omega=0$~MeV to $\omega=300$~MeV.


In Sect.~\ref{sec:results} we discuss our predictions for the deuteron Compton 
cross sections for three different energies
between 50~MeV and 100~MeV, comparing to experimental data and to the 
$\mathcal{O}(q^3)$ 
HB$\chi$PT calculation~\cite{Phillips}. Before that, we give a brief survey of
the theoretical formalism in Sect.~\ref{sec:theory}. There we show that 
combining Weinberg's counting ideas with the SSE power counting scheme leads 
to no additional diagrams in the two-body part of the kernel with respect 
to~\cite{Phillips}. 
In Sect.~\ref{sec:fits} we present our results for the isoscalar
polarizabilities, derived from a fit to 
elastic deuteron Compton scattering data, which turn out to be in good 
agreement with the theoretical expectation that the isovector 
components are small. We conclude in Sect.~\ref{sec:outlook} 
and give a brief outlook on future projects, one of which aims to cure the 
shortcomings of our calculation in the extreme low-energy regime 
$\omega\ll50$~MeV (cf. Sect.~\ref{sec:theory}).
 
\section{Compton Scattering off the Deuteron in Effective Field Theory}
\label{sec:theory}
We are calculating Compton scattering off the deuteron in the framework of the
Small Scale Expansion~\cite{HHKLett}, an Effective Field Theory with nucleons,
pions and the 
$\Delta(1232)$ resonance as explicit degrees of freedom. 
In this extension of $\chi$PT, the expansion parameter is called $\epsilon$, 
denoting either a small momentum, the pion mass or
the mass difference $\Delta_0$ between the real part of the 
$\Delta$ mass and the nucleon mass. 
The relevant pieces of the $N\pi$ and $\Delta\pi$ 
Lagrangeans have been discussed in the literature many times, and we refer 
the interested reader to~\cite{BKM} for the $N\pi$ Lagrangean and for a 
general review of HB$\chi$PT, 
and to~\cite{HHKK} and~\cite{HGHP} for the relevant pieces of the 
$\Delta\pi$ Lagrangean.

The power-counting scheme that we use for Compton scattering off light nuclei 
is motivated by Weinberg's idea to 
count powers only in the interaction kernel. 
We base our calculation on the hybrid approach, which is a well-established 
tool by now. 
While the kernel is 
power counted according to the rules of the Effective Field Theory, the 
deuteron wave functions we use are obtained from 
state-of-the-art $NN$ potentials: Nijm93~\cite{Nijm}, the 
CD-Bonn potential~\cite{Bonn}, the AV18 potential~\cite{AV18} and the NNLO 
chiral potential~\cite{Epelbaum}, where this last potential also follows 
Weinberg's suggestion, and is derived by applying HB$\chi$PT power counting to 
the $N N$ potential $V$.

It is convenient to write the Green's function for Compton scattering from 
the two-nucleon system as
\beq
G_{\gamma\gamma}=G\,K_{\gamma\gamma}\,G+G\,K_\gamma\,G\,K_\gamma\,G\,,
\eeq
with $K_{\gamma\gamma}$ the 
two-nucleon-irreducible part of the interaction kernel, which contains both 
one-body and two-body mechanisms, and $K_\gamma\,G\,K_\gamma$ the so-called 
nuclear-resonance contribution\footnote{The 
nomenclature is due to nuclear resonances which are excited by the initial 
interaction with the photon and which one might expect to dominate the 
Compton process at low energies, see e.g.~\cite{Lvov}.} to the kernel. 
$G$ is the two-particle Green's function, constructed from the 
two-nucleon-irreducible interaction $V$ and the free two-nucleon Green's 
function. We apply the same power counting rules to both $K_{\gamma\gamma}$ 
and $K_\gamma G K_\gamma$, calculating all contributions up to a specific 
order in $\epsilon$, namely $\epsilon^3$. 

To do this we first note that a diagram contributing at a certain 
order in $q$ in HB$\chi$PT contributes at the same order $\epsilon$ in SSE.
However, there are some kinematics in which HB$\chi$PT counting should not be 
employed for the $\gamma N N\rightarrow \gamma N N$ kernel.
In HB$\chi$PT the leading order 
propagator of a nucleon with the energy $\omega$ of the external probe 
flowing through it is 
$\frac{i}{\omega}$. Corrections from the kinetic energy of the nucleon are
treated perturbatively. In the deuteron, such a perturbative treatment is not 
applicable for low photon energies, due to the relative momentum $\vec{p}$ 
of the two nucleons. Therefore, in the low-energy regime one has to use 
the full non-relativistic nucleon propagator $\frac{i}{\omega-p^2/2M}$. 
Nonetheless,
the approximation  $\frac{i}{\omega}$ is useful for $\omega\gg B$, with 
$B\approx 2.225$~MeV the binding energy of the deuteron, as $p^2/M\sim B$ with 
a typical nucleon momentum $p$ inside the deuteron. 
These considerations demonstrate that for $\omega\gg B$ the nucleon 
propagator may be counted as $\mathcal{O}(\epsilon^{-1})$  
like in standard HB$\chi$PT, whereas in the `nuclear' regime 
$\omega\sim \mathcal{O}(m_\pi^2/M)$ 
it has to be counted as $\mathcal{O}(\epsilon^{-2})$, as $p\sim\epsilon$.
Therefore, one has to strictly differentiate between two energy regimes: 
the nuclear regime $\omega\sim \mathcal{O}(m_\pi^2/M)$  
and the regime $\omega\sim \mathcal{O}(m_\pi)$.
Here we work in the latter one, as we are mainly concerned with photon 
energies $\omega\geq50$~MeV, which is the energy region where one starts to be
sensitive to the nucleon polarizabilities.

We note further that only in the regime $\omega\sim \mathcal{O}(m_\pi)$ can 
one treat 
the contributions from $K_\gamma G K_\gamma$ using a perturbative chiral 
expansion. Since we do treat this piece using HB$\chi$PT counting it is 
therefore no surprise that the Compton low-energy theorems are violated. For 
example, the Thomson limit for Compton scattering from a nucleus of charge 
$Z\,e$ and mass $A\,M$,
\beq
A_\mathrm{Thomson}=A(\omega=0)=
       -\frac{Z^2\,e^2}{A\,M}\,\vec{\epsilon}\cdot\vec{\epsilon}\,'\,,
\eeq
a direct consequence of gauge invariance \cite{Friar}, cannot be recovered
without the full resonance term. 
Therefore, we strictly constrain ourselves to photon energies 
$\omega\sim \mathcal{O}(m_\pi)$, where a 
perturbative expansion of the kernel in the standard HB$\chi$PT counting 
scheme, 
i.e. counting the nucleon propagator as $\mathcal{O}(\epsilon^{-1})$, 
is possible. The lower limit of this power counting  turns out to be 
$\omega\approx 50$~MeV, so we have to caution the reader that the 
calculation is  not supposed to work in the region  $\omega\ll 50$~MeV
(an extension to lower energies is in progress~\cite{future}).
A more detailed discussion of the power counting applied to the meson exchange
part of our calculation can be found in Refs.~\cite{Phillips,McGPhil}. 
For an Effective Field Theory approach to deuteron Compton scattering where 
pions are integrated out, see Ref.~\cite{Rupak}. These calculations describe 
the very-low-energy region well and also reach the exact Thomson limit.

The $T$-matrix for Compton scattering off the deuteron is 
derived as the matrix element of the interaction kernel,
evaluated between an 
initial and final state deuteron wave function, as explained in great detail 
in~\cite{Phillips},
\beq
T=\left<\Psi_f|K_{\gamma\gamma}+K_\gamma\,G\,K_\gamma|\Psi_i\right>\,.
\eeq
Stated differently, one obtains $T$ by extracting the piece of 
$G_{\gamma\gamma}$ corresponding 
to the deuteron pole at $E=-B$ in both the initial and final state.

\begin{figure}[!htb]
\begin{center} 
\includegraphics*[width=.3\textwidth]
{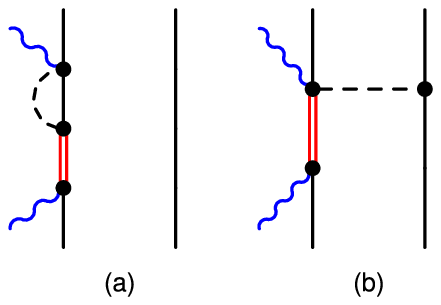}
\parbox{1.\textwidth}{
\caption{Two examples of $\mathcal{O}(\epsilon^4)$ contributions to Compton 
scattering on the deuteron including explicit $\Delta(1232)$ degrees of 
freedom for one- and two-body contributions.}
\label{SSEOQ4}}
\end{center}
\end{figure}

As we are calculating $\gamma d$ scattering in the Small Scale Expansion, we 
also have to fix our 
counting rules for diagrams including $\Delta(1232)$ propagators. For the 
one-body contributions this is straightforward, as we apply the 
SSE counting scheme, cf. Refs.~\cite{HGHP, HHK}. 
As far as the two-body physics is concerned,
we combine the SSE counting rules, e.g. counting the $\Delta$-propagator as 
$\epsilon^{-1}$, with Weinberg's prescription of counting only within the 
interaction kernel. To $\mathcal{O}(\epsilon^3)$, the order up to which we are
working, this leads to identical meson exchange diagrams as in the 
$\mathcal{O}(q^3)$ HB$\chi$PT calculation. All additional diagrams are at 
least one order higher, an example is given in Fig.~\ref{SSEOQ4}(b) (an 
example of an $\mathcal{O}(\epsilon^4)$ one-body diagram is sketched 
in Fig.~\ref{SSEOQ4}(a)). A modified counting 
scheme in the two-body sector has been suggested in~\cite{Beane}, as certain
pion-exchange diagrams may be enhanced when the photon energy comes close 
enough to the pion mass that the pions in the two-body diagrams are almost on 
mass shell. We do not consider such a modification necessary for 
our calculation, as we restrict ourselves to photon energies 
$\omega\leq100$~MeV.

Therefore, the diagrams contributing to deuteron Compton scattering up to 
$\mathcal{O}(\epsilon^3)$ are:
\begin{itemize}
\item One-body contributions without explicit $\Delta(1232)$ degrees of 
freedom. These are the single-nucleon seagull with the two-photon vertex from
$\mathcal{L}_{N\pi}^{(2)}$~(Fig.~\ref{chiPTsingle}(a)), which is the only 
$\mathcal{O}(\epsilon^2)$ contribution, 
the nucleon-pole terms (Fig.~\ref{chiPTsingle}(b)), the 
pion pole (Fig.~\ref{chiPTsingle}(c)) and the contributions from the leading 
chiral dynamics of the pion cloud around the nucleon 
(Figs.~\ref{chiPTsingle}(d)-(g)). Up to third order, the only difference in 
these diagrams, 
compared to Compton scattering off the single nucleon, is that the pole 
diagrams (Fig.~\ref{chiPTsingle}(b)), which are conveniently calculated 
in the $\gamma N$ center of mass frame, have to be boosted 
to the $\gamma N N$ center of mass system, as the calculation is performed in 
the $\gamma d$ cm frame. The resulting formulae from the boost are given 
in~\cite{Phillips}.
\begin{figure}[!htb]
\begin{center} 
\includegraphics*[width=.121\linewidth]
{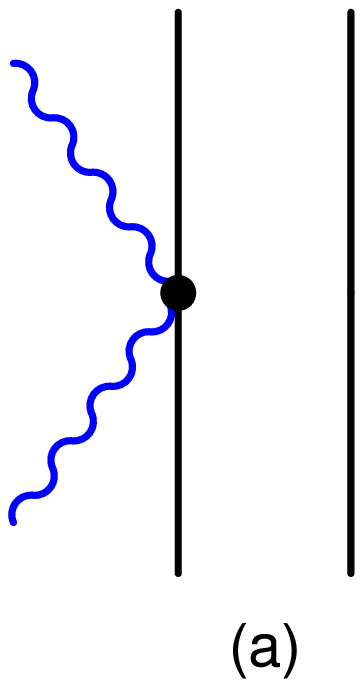}
\includegraphics*[width=.121\linewidth]
{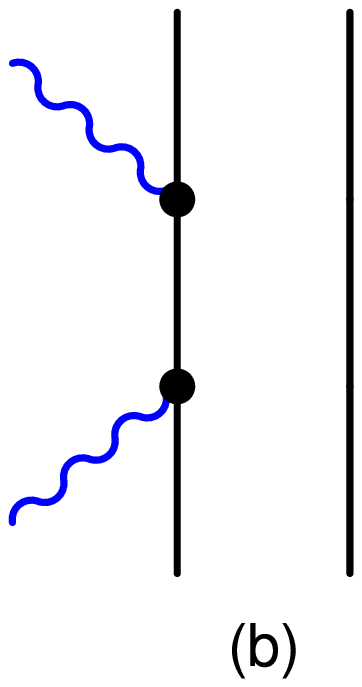}
\includegraphics*[width=.121\linewidth]
{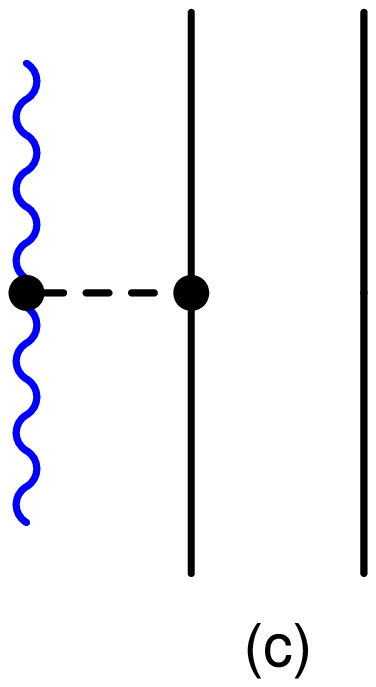}
\includegraphics*[width=.121\linewidth]
{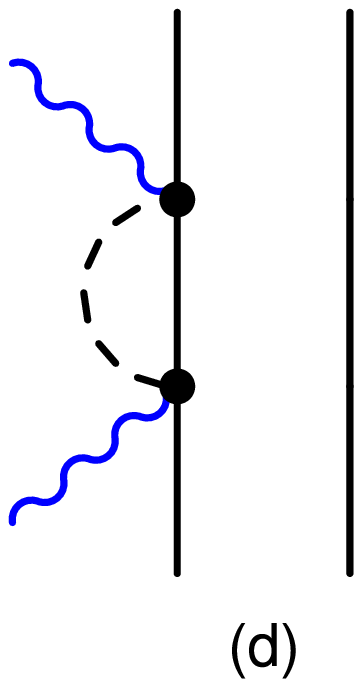}
\includegraphics*[width=.121\linewidth]
{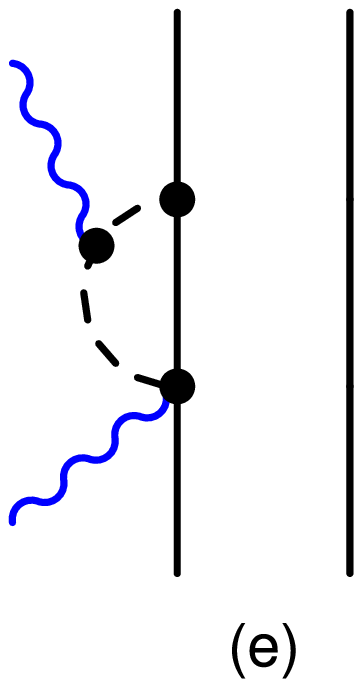}
\includegraphics*[width=.121\linewidth]
{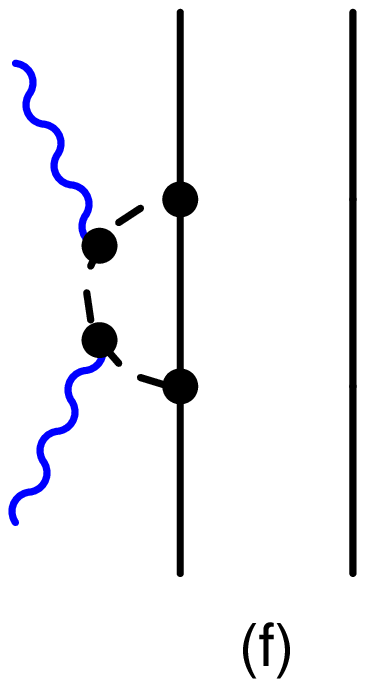}
\includegraphics*[width=.121\linewidth]
{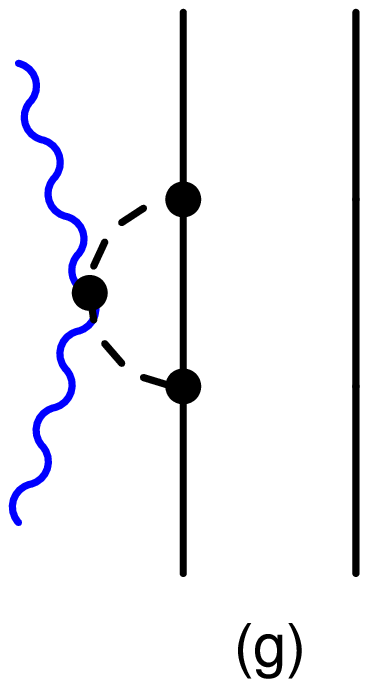}
\parbox{1.\textwidth}{
\caption{One-body interactions without a $\Delta(1232)$ propagator 
contributing to Compton scattering on the 
deuteron up to $\mathcal{O}(\epsilon^3)$ in SSE. Permutations and crossed 
graphs are not shown.}
\label{chiPTsingle}}
\end{center}
\end{figure}
\item One-body diagrams with explicit $\Delta$ degrees of freedom, as 
shown in Fig.~\ref{SSEsingle}: The $\Delta$-pole diagrams 
(Fig.~\ref{SSEsingle}(a)) and the contributions from the pion cloud around 
the $\Delta(1232)$ (Figs.~\ref{SSEsingle}(b)-(e)). 
\item Two isoscalar short-distance one-body 
operators (Fig.~\ref{SSEsingle}(f)), which 
give energy-inde-pendent contributions to the dipole polarizabilities 
$\alpha_E^s$ and $\beta_M^s$. 
They are formally of $\mathcal{O}(\epsilon^4)$ but 
turn out to give an anomalously large contribution to the single-nucleon 
Compton amplitude. Therefore, they have to be promoted to next-to-leading 
order as discussed in detail in~\cite{HGHP}, which we also refer to for the 
Lagrangean. We note that 
except for the two 
contact operators (Fig.~\ref{SSEsingle}(f)), the $\delta$-expansion 
(cf. Sect.~\ref{sec:introduction}) up to NNLO 
is equivalent to $\mathcal{O}(\epsilon^3)$ SSE in the energy range 
$\omega\sim m_\pi$ considered. 
\begin{figure}[!htb]
\begin{center} 
\includegraphics*[width=.121\linewidth]
{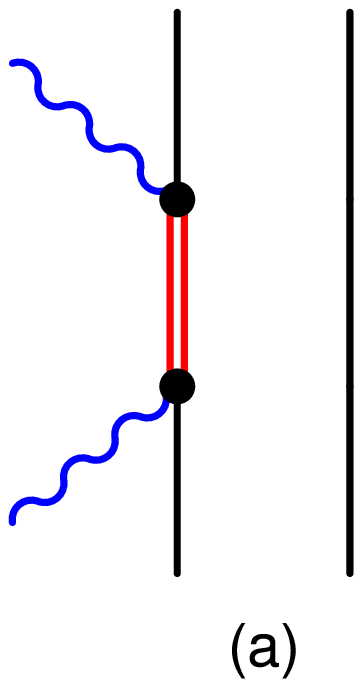}
\includegraphics*[width=.121\linewidth]
{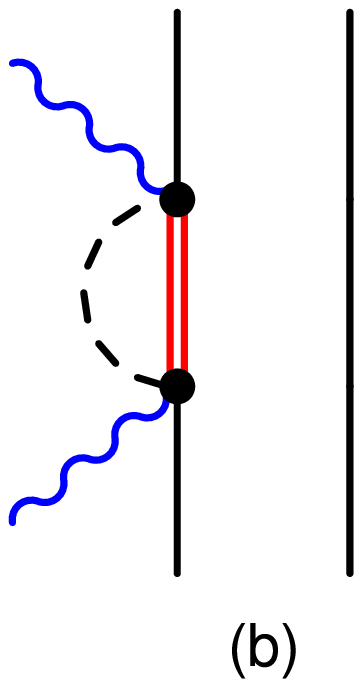}
\includegraphics*[width=.121\linewidth]
{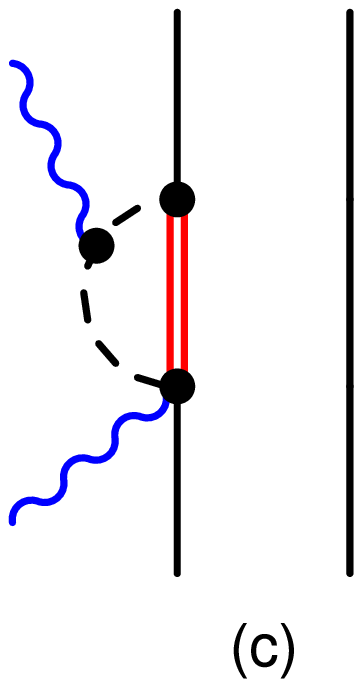}
\includegraphics*[width=.121\linewidth]
{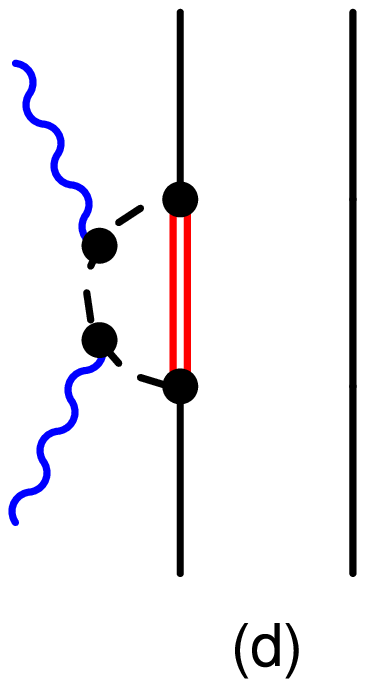}
\includegraphics*[width=.121\linewidth]
{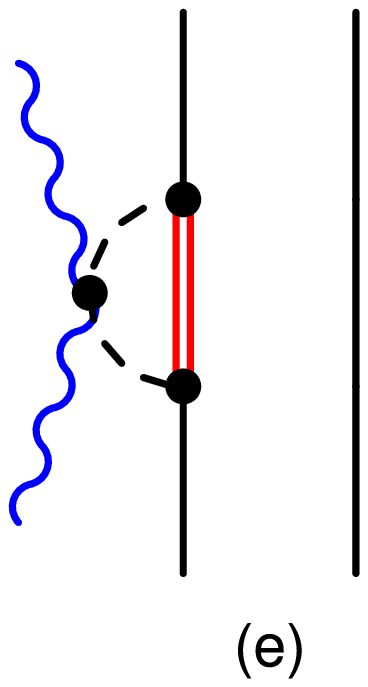}
\includegraphics*[width=.121\linewidth]
{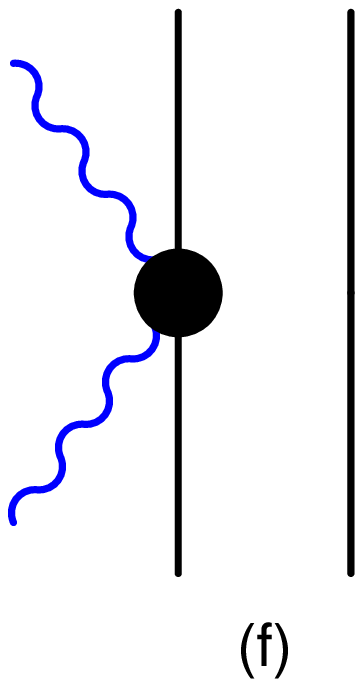}
\parbox{1.\textwidth}{
\caption{One-body interactions which contribute to deuteron Compton 
scattering at $\mathcal{O}(\epsilon^3)$ in SSE in addition 
compared to third-order HB$\chi$PT. Permutations and crossed graphs 
are not shown.}
\label{SSEsingle}}
\end{center}
\end{figure}
\item Two-body contributions with one pion exchanged between the two 
nucleons (Fig.~\ref{chiPTdouble}). 
In total there 
are nine two-body diagrams at $\mathcal{O}(\epsilon^3)$. As discussed 
before, the meson exchange diagrams are identical in third-order HB$\chi$PT 
and SSE.
\begin{figure}[!htb]
\begin{center} 
\includegraphics*[width=.75\textwidth]
{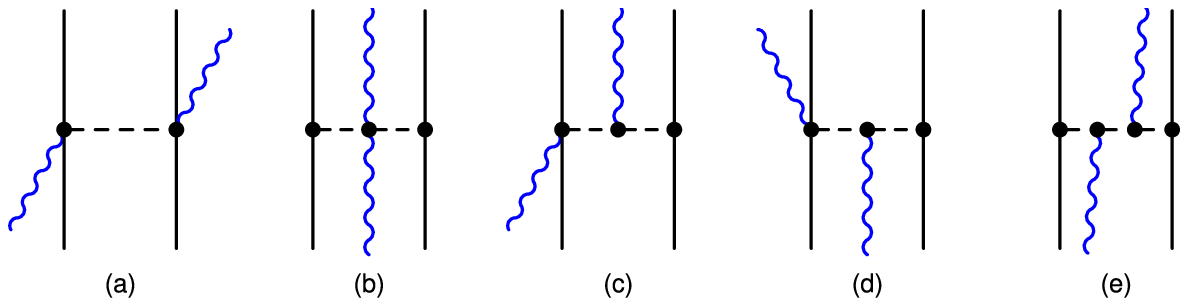}
\parbox{1.\textwidth}{
\caption{Two-body interactions 
contributing to the kernel for Compton scattering on the 
deuteron at $\mathcal{O}(\epsilon^3)$ in SSE. Diagrams which differ only by 
nucleon interchange are not shown.}
\label{chiPTdouble}}
\end{center}
\end{figure}
\end{itemize}

All these diagrams (Figs.~\ref{chiPTsingle}--\ref{chiPTdouble}) make up our 
interaction kernel. The SSE single-nucleon amplitudes can 
be found in~\cite{HGHP}, while the two-body contributions are given 
explicitly in \cite{Phillips}. 
Note that we have simplified the expressions given in~\cite{HGHP} with respect
to the exact position of the pion threshold as we are only analysing Compton 
scattering for photon energies $\leq 100$~MeV.
An estimate of the (small) size of this simplification is given in 
Sect.~\ref{sec:threshold}. 

In the next section we compare our $\mathcal{O}(\epsilon^3)$ SSE
results for the deuteron Compton cross 
sections to the $\mathcal{O}(q^3)$ HB$\chi$PT calculation performed 
in~\cite{Phillips} and to the 
available experimental data. Special interest is put on the energy and 
wave-function dependence of the cross section.

\section{Predictions for Deuteron Compton Cross Sections}
\label{sec:results}

\begin{figure}[!htb]
\begin{center} 
\includegraphics*[width=.48\textwidth]{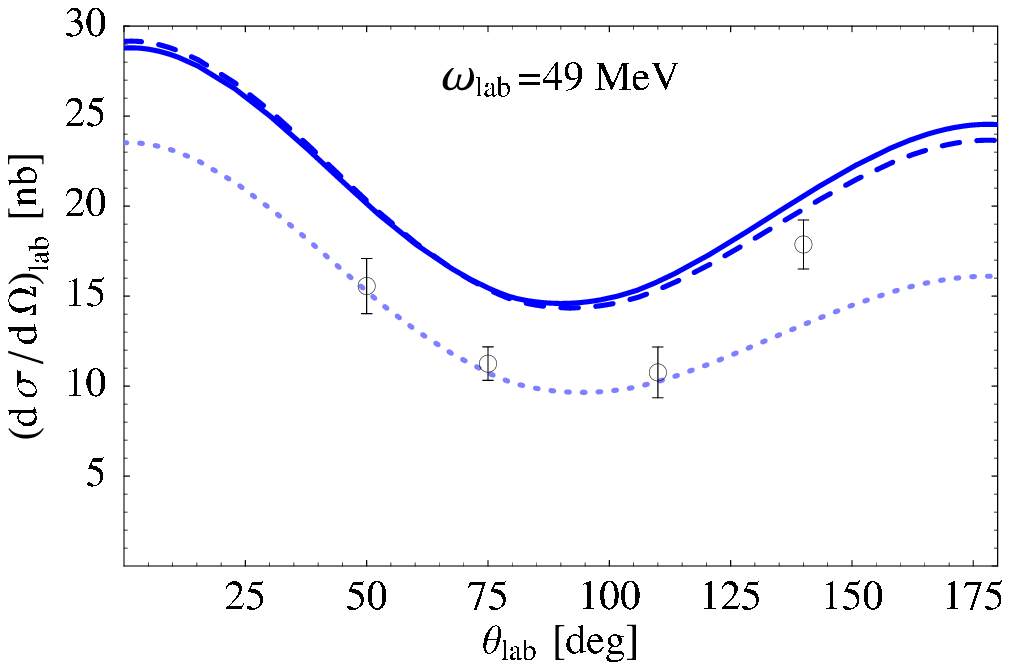}
\hfill
\includegraphics*[width=.48\textwidth]{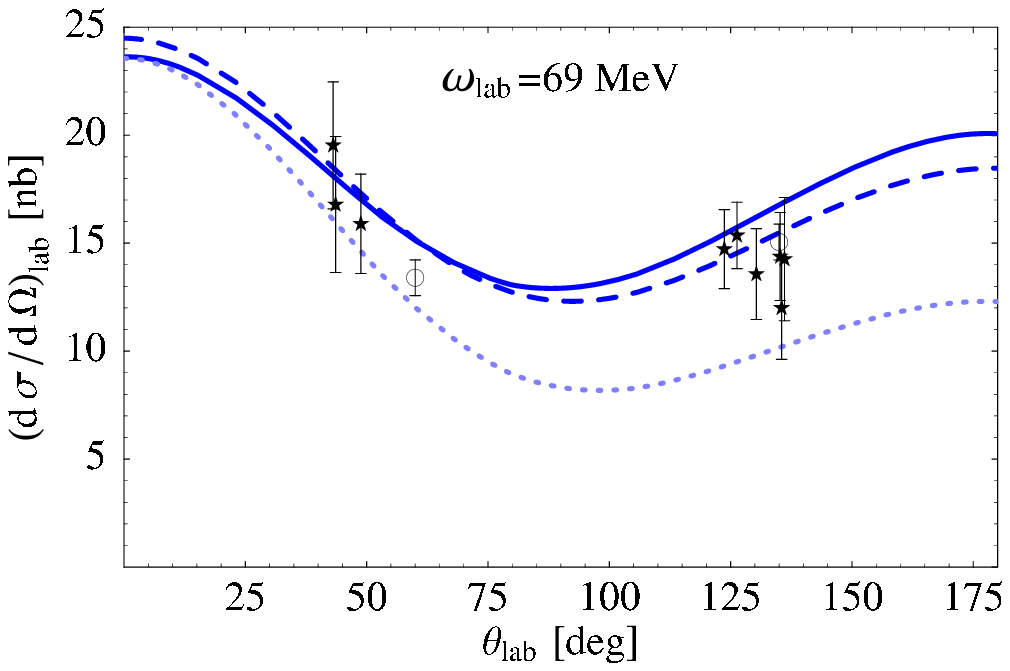}
\includegraphics*[width=.6\textwidth]{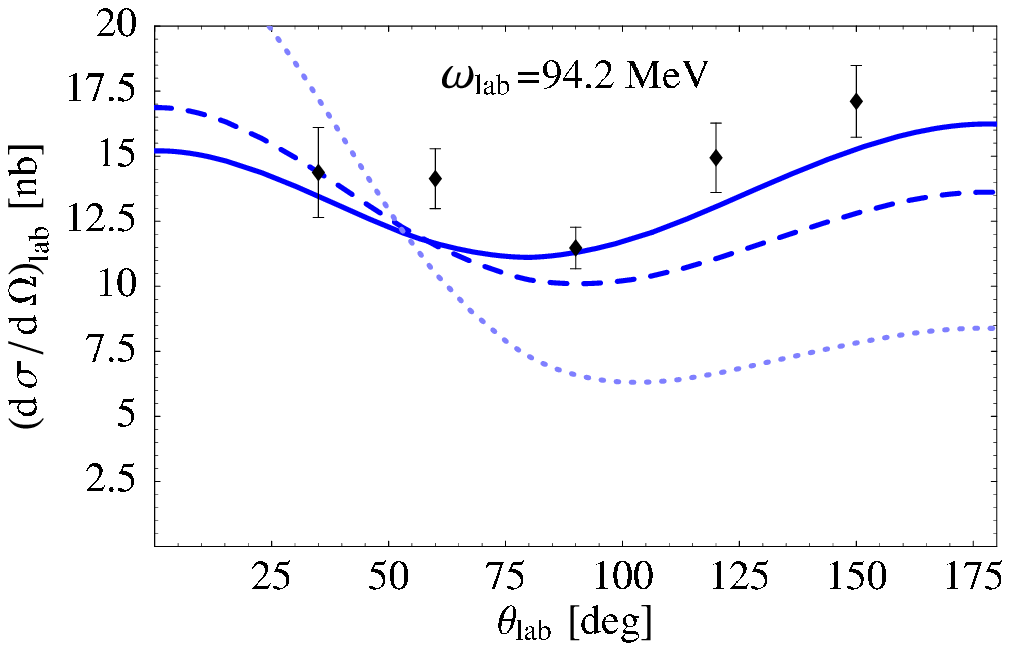}
\parbox{1.\textwidth}{
\caption{Comparison of the $\mathcal{O}(q^3)$ HB$\chi$PT (dashed) and the 
$\mathcal{O}(\epsilon^3)$ SSE (solid) prediction at 
$\omega_\mathrm{lab}=49\;\mathrm{MeV}$, $\omega_\mathrm{lab}=69\;\mathrm{MeV}$
and $\omega_\mathrm{lab}=94.2\;\mathrm{MeV}$ using the chiral NNLO wave 
function~\cite{Epelbaum}. The data are from Illinois~\cite{Lucas} (circle), 
Lund~\cite{Lund} (star) and SAL~\cite{Hornidge} (diamond). The dotted line is 
the $\mathcal{O}(q^2)$ result.}
\label{SSEHBplots}}
\end{center}
\end{figure}

In Fig.~\ref{SSEHBplots} we compare the $\mathcal{O}(\epsilon^3)$ SSE 
predictions 
to the $\mathcal{O}(q^3)$ HB$\chi$PT calculation of Ref.~\cite{Phillips}, using
the wave function derived from the NNLO chiral potential with spectral 
function regularization~\cite{Epelbaum}. 
We also show the $\mathcal{O}(q^2)$ 
result, which consists only of the single-nucleon seagull 
(Fig.~\ref{chiPTsingle}(a)). The 
experiments shown have been performed at a lab-energy of 49~MeV~\cite{Lucas}, 
$\sim$67~MeV~\cite{Lund}, 
69~MeV~\cite{Lucas} and $\sim$94.2~MeV~\cite{Hornidge}. (The last experiment
used photons in an energy range from $84.2-104.5$~MeV; the deviation from the
central value has been corrected for~\cite{Hornidge}.) 

The values we use for physical constants can be found in Table~\ref{tab:const}.
 For the coupling constants connected with the two short-distance 
$\gamma N$-operators 
(cf.~Sect.~\ref{sec:theory}) and the $\gamma N\Delta$ coupling $b_1$ 
we use the results of the Baldin Sum Rule constrained fit to the spin-averaged 
proton Compton scattering data from~\cite{HGHP}. 
Fitting the short-distance operators is equivalent to fitting the static 
polarizabilities $\alpha_E^p$, $\beta_M^p$. 
We use the central values of the fit, which are 
$\alpha_E^p=11.04\cdot10^{-4}\;\mathrm{fm}^3$, 
$\beta _M^p= 2.76\cdot10^{-4}\;\mathrm{fm}^3$~\cite{HGHP}, cf. 
Sect.~\ref{sec:introduction}. 
The $\mathcal{O}(\epsilon^3)$ SSE calculation then predicts 
$\alpha_E^n\equiv\alpha_E^p$, $\beta _M^n\equiv\beta _M^p$ 
as the isovector contributions only come in at $\mathcal{O}(\epsilon^4)$.
Therefore, there are no free parameters in our deuteron Compton calculation.

\begin{table}[!htb] 
\begin{center}
\begin{tabular}{|c|c|c|}
\hline 
Parameter & Value & Comment \\
\hline 
$m_\pi$  & $139.6$ MeV & charged pion mass \\
$M$ & $938.9$ MeV & isoscalar nucleon mass \\
$f_\pi$  & $92.4$ MeV & pion decay constant \\
$g_A$ & $1.267$& axial coupling constant \\
$\alpha$ & $1/137$& QED fine structure constant \\
$\kappa_v$ & $3.708$ & isovector anomalous magnetic moment  \\
$\kappa_s$ & $-0.118$ & isoscalar anomalous magnetic moment \\
\hline
$M_d$&$1875.58$ MeV& deuteron mass \\
$B$&$2.2246$ MeV& deuteron binding energy\\
\hline
$\Delta_0$  & $271.1$ MeV & $N\Delta$ mass splitting \\
$g_{\pi N\Delta}$& $1.125$& $\pi N\Delta$ coupling constant \\
$b_1$&$4.67$& $\gamma N\Delta$ coupling constant\\
$g_{117}$&$18.82$& short-distance coupling constant\\
$g_{118}$&$-6.05$& short-distance coupling constant\\
\hline
\end{tabular}
\end{center}
\caption{$\chi$EFT parameters determined independently of deuteron 
Compton scattering. 
Magnetic moments are given in nuclear magnetons.}
\label{tab:const}
\end{table}

From the 49 and 69~MeV curves shown in Fig.~\ref{SSEHBplots} it is obvious 
that explicit $\Delta$
degrees of freedom may well be neglected for these low energies. The two 
calculations -- HB$\chi$PT and SSE -- yield results which differ only within 
the uncertainties one expects from higher order contributions. 
This is an important check, as it demonstrates the correct decoupling of the 
resonance, which provides the same low-energy limit in both 
theories.

Another interesting observation can be made: The counting scheme described in 
Sect.~\ref{sec:theory} seems to break down for energies somewhere near 50~MeV,
as both theoretical descriptions miss the 49~MeV 
data points, whereas the 69~MeV data (we neglect the minor corrections due to 
the data of~\cite{Lund} being measured around 67~MeV) are well described 
within both theories. The 49~MeV data are best described by the 
$\mathcal{O}(q^2)$ calculation but we believe this is a coincidence, as 
the low-energy theorems are violated at this order too.
However, for higher energies -- i.e. for describing the 
94.2~MeV data correctly -- the inclusion of the explicit $\Delta$ field seems 
to be advantageous in a third-order calculation.
Here, $\mathcal{O}(q^3)$ $\chi$PT  
misses the data in the backward direction. It also fails to reproduce the 
shape of 
the data points, which shows a slight tendency towards higher cross sections 
in the backward than in the forward direction. This shape is very well 
reproduced in SSE, demonstrating once again the importance of the $\Delta$ 
resonance in Compton backscattering, due to the strong $M1\rightarrow M1$ 
transition. We note that this feature 
can be clearly seen in the dynamical magnetic dipole polarizability 
$\beta_{M1}(\omega)$, even for photon energies below the pion-production 
threshold (cf.~\cite{HGHP} and Sect.~\ref{sec:unbiased}). Calculations
like the ones presented in Refs.~\cite{Lvov,Miller}, which include the 
dynamics of the polarizabilities 
only via the leading~\cite{Lvov,Miller} and subleading terms~\cite{Lvov} of a 
Taylor expansion, may therefore fail to describe the data around 95~MeV.

\subsection{Energy Dependence of the $\gamma d$ Cross Sections}

\begin{figure}[!htb]
\begin{center} 
\includegraphics*[width=.48\textwidth]{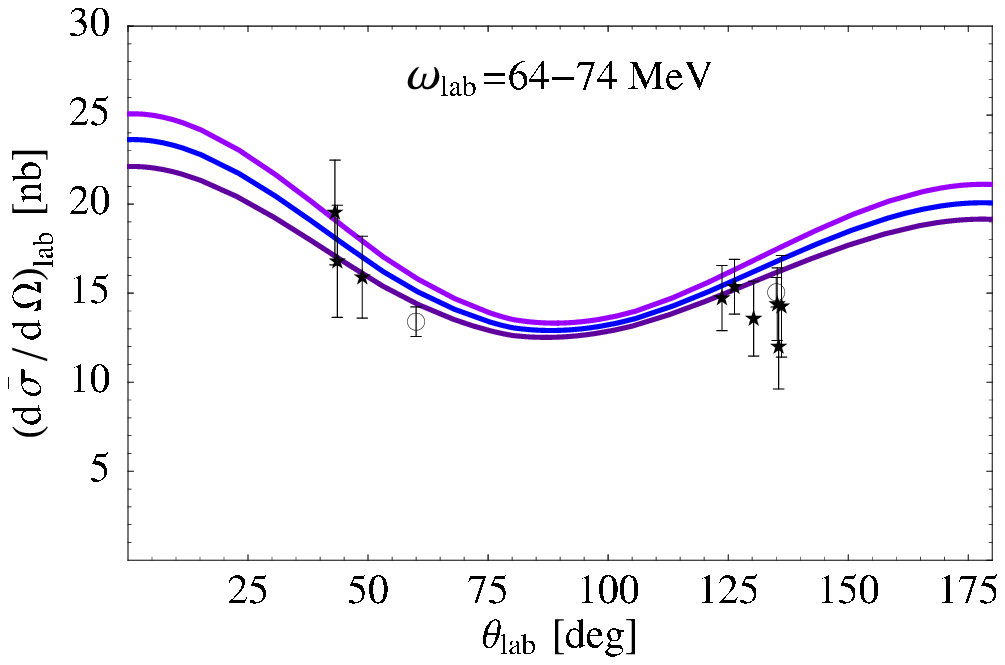}
\hfill
\includegraphics*[width=.48\textwidth]{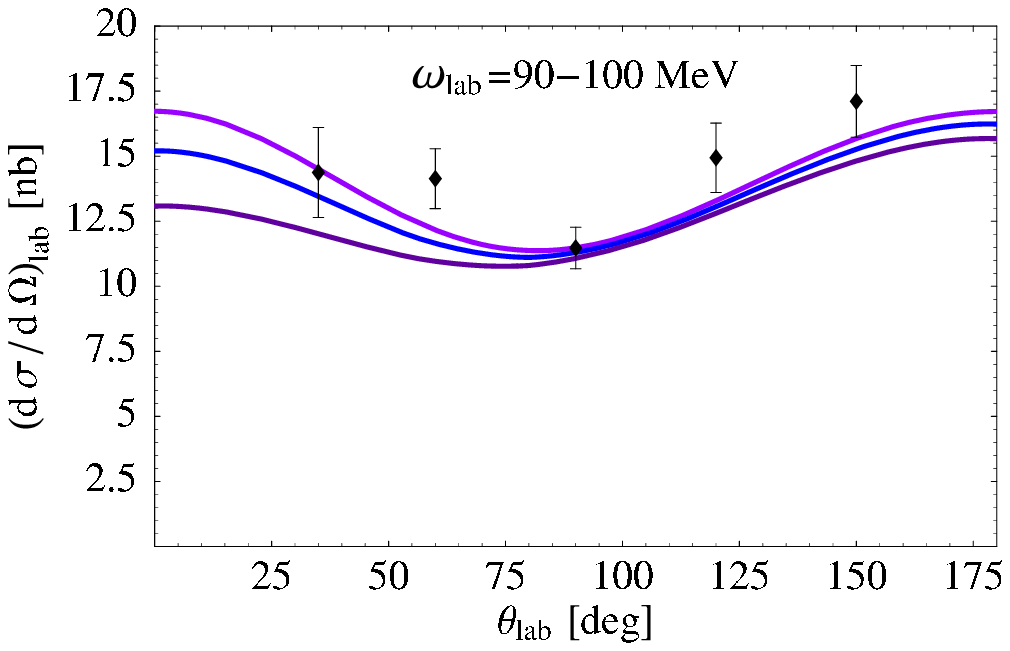}
\parbox{1.\textwidth}{
\caption{$\mathcal{O}(\epsilon^3)$ SSE results for 64~MeV, 69~MeV, 74~MeV and,
respectively,
90~MeV, 94.2~MeV, 100~MeV (from the upper to the lower curve in each panel), 
using the $\chi$PT wave function~\cite{Epelbaum}. } 
\label{omegadep}}
\end{center}
\end{figure}

In order to decrease the statistical uncertainties, the 
experiment~\cite{Hornidge} had to accept 
scattering events in an energy range of around 20~MeV. Therefore we think it 
worthwhile to examine the sensitivity of our results to the photon energy.
In fact, our calculations suggest that the forward-angle cross section, in 
particular, has a sizeable energy dependence, which is, however, nearly 
linear. In 
Fig.~\ref{omegadep} we show our results for three different photon energies
around 69~MeV and 94.2~MeV, respectively,
separated only by 10 MeV. This emphasizes the importance of having a 
well-defined photon energy at which to examine the effects of $\alpha_E$ and
$\beta_M$.

\subsection{Correction due to the Pion-Production Threshold}
\label{sec:threshold}

\begin{figure}[!htb]
\begin{center} 
\includegraphics*[width=.5\textwidth]{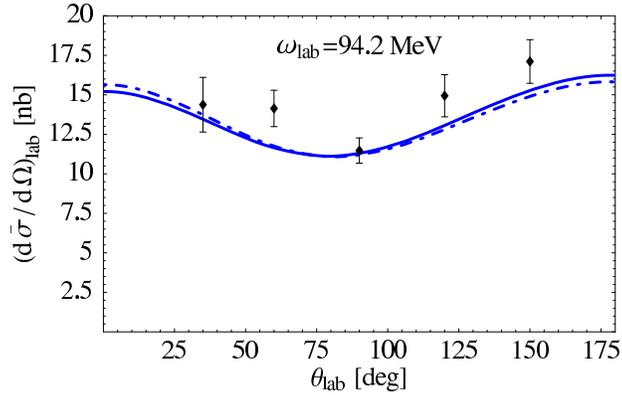}
\parbox{1.\textwidth}{
\caption{Estimate of the effect of a threshold correction (dotdashed) on the 
$\mathcal{O}(\epsilon^3)$ SSE results 
(solid), using the chiral NNLO wave function~\cite{Epelbaum}.}
\label{SSEnsuns}}
\end{center}
\end{figure}

In low-order HB$\chi$PT/SSE calculations the $\gamma d\rightarrow \pi N N$
threshold is not at the 
correct position as dictated by relativistic kinematics. For a similar problem,
regarding the correct position of the pion-production threshold in the 
single-nucleon sector, see e.g.~\cite{HGHP}. So far we refrain from 
an analogous correction for $\gamma d$ scattering.
However, in Fig.~\ref{SSEnsuns} we investigate what deviations one would 
expect from our present results, as indicated by an estimate, where we 
use the single-nucleon SSE amplitudes~\cite{HGHP} with the exact expression 
for $\sqrt{s}-M$. Obviously, even at the highest photon energies considered 
here, 94.2~MeV, the corrections are negligible, given 
the sizeable error bars of the experimental data and the theoretical 
uncertainties of a leading-one-loop order calculation.

\subsection{Wave-Function Dependence of the $\gamma d$ Cross Sections}
\label{wfdep}

\begin{figure}[!htb]
\begin{center} 
\includegraphics*[width=.48\textwidth]{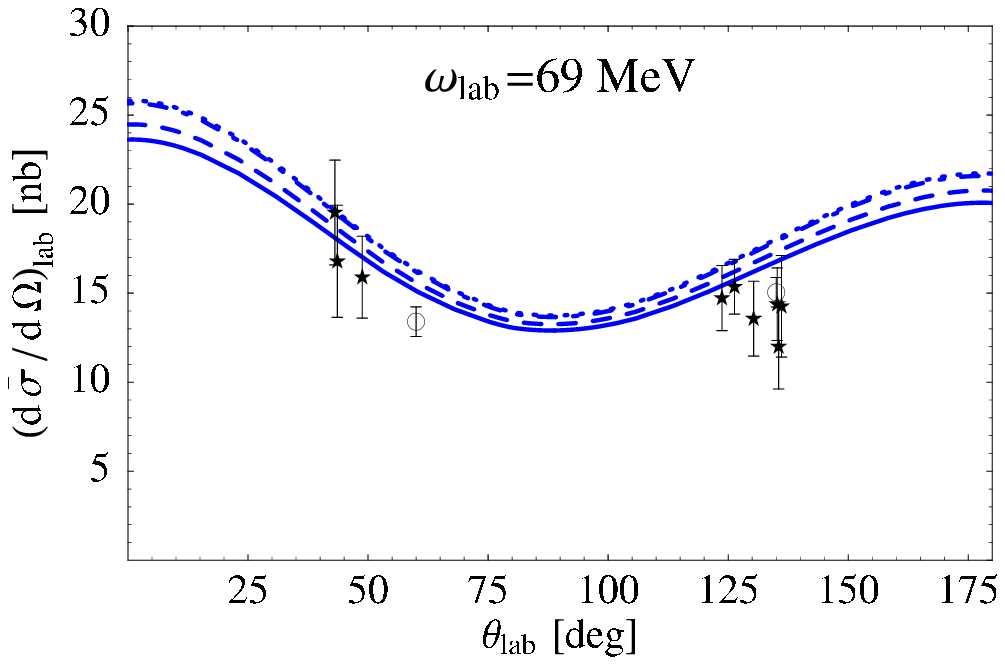}
\hfill
\includegraphics*[width=.48\textwidth]{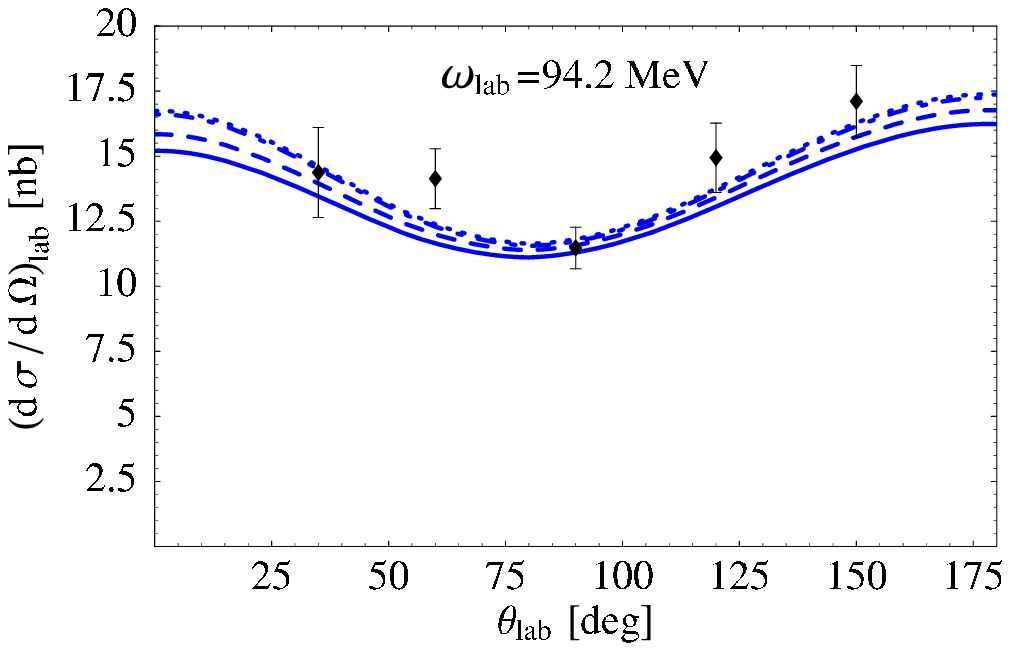}
\parbox{1.\textwidth}{
\caption{
$\mathcal{O}(\epsilon^3)$ SSE results for 69 and 94.2~MeV, using four
different wave functions: NNLO $\chi$PT~\cite{Epelbaum} (solid), 
Nijm93~\cite{Nijm} (dotted), 
CD-Bonn~\cite{Bonn} (dashed), AV18~\cite{AV18} (dotdashed).} 
\label{wavefdep}}
\end{center}
\end{figure}

Another interesting issue is the wave-function dependence of our results.
Fig.~\ref{wavefdep} investigates the sensitivity to the 
wave function chosen, showing sizeable deviations between the NNLO $\chi$PT
wave function~\cite{Epelbaum} 
on one hand and the  wave functions derived from the Nijm93 
potential~\cite{Nijm} and the AV18 potential~\cite{AV18} on the other. 
The last two yield results which  are nearly indistinguishable but 
are considerably higher than those found with the wave function 
of~\cite{Epelbaum}. 
With the CD-Bonn wave function~\cite{Bonn} we obtain results in between 
NNLO $\chi$PT and Nijm93/AV18.
This pattern is identical for both energies under investigation, 69~MeV and 
94.2~MeV. 
Given that our calculation is based on a low-energy Effective Field Theory of 
QCD, the dependence on the wave function is somewhat worrisome and will 
be discussed further in~\cite{future}. According to Weinberg counting it is an 
$\mathcal{O}(q^5)$ effect, so a deviation of the order of 10\% is more than 
one would expect.
We interpret this feature as an unwanted sensitivity to short-distance 
physics, because the long-range part of all wave functions, described by  
one- and two-pion exchange, is identical. However, one must caution that the 
NNLO $\chi$PT potential reproduces the Nijmegen partial-wave analysis 
with less precision than the CD-Bonn, AV18 or Nijm93 potentials. 

In this section we presented our predictions for $\gamma d$ differential 
cross sections. These are parameter-free as we fixed the nucleon 
polarizabilities via proton Compton data. 
The good agreement of the SSE results with experiment at 69~MeV and 94.2~MeV 
leaves little room for large isovector polarizabilities, since these 
predictions used the same values for the proton and neutron polarizabilities.
It further 
encourages us to determine the isoscalar dipole polarizabilities $\alpha_E^s$ 
and $\beta_M^s$ directly from the deuteron Compton cross sections. 
The results are displayed in the next section, together 
with the results one obtains from analogous 
fits using the $\mathcal{O}(q^3)$~HB$\chi$PT amplitudes.

\section{Determining $\alpha_E^s$ and $\beta_M^s$ from $\gamma d$ 
Scattering}
\label{sec:fits}

An accurate and systematically-improvable description of Compton scattering 
on deuterium offers the 
possibility to extract the isoscalar polarizabilities directly 
from deuteron Compton scattering experiments in a systematic way. The 
resulting numbers can then be 
combined with the known numbers for the proton to draw 
conclusions about isovector pieces $\alpha_E^v$ and $\beta_M^v$, or, 
equivalently, the elusive neutron polarizabilities.
As our SSE 
calculation provides a reasonable description of the 69~MeV and the 94.2~MeV 
data (see Sect.~\ref{sec:results}), we present in the following our results 
from a fit of the isoscalar polarizabilities to these two data sets. This 
corresponds to fitting the coupling strengths of the two short-distance 
isoscalar $\gamma N$-operators (Fig.~\ref{SSEsingle}(f)), which we now fit to 
$\gamma d$ data rather than to $\gamma p$ data. In this way we can check 
our assumption that the short-distance operators are isoscalar at leading 
order. If the value extracted from $\gamma d$ data is approximately that from 
$\gamma p$ data, that argues in favour of short-distance mechanisms which 
are predominantly isoscalar. 

Our SSE results are compared to the fit
results that we get for $\alpha_E^s$ and $\beta_M^s$ when we use modified 
$\mathcal{O}(q^3)$~HB$\chi$PT amplitudes. This modification consists of 
including in our calculation isoscalar short-distance $\gamma N$ 
operators which change 
both the electric and magnetic polarizability from their $\mathcal{O}(q^3)$ 
values. In other words, we write
\begin{align}
\alpha_E^s&=\frac{5\,\alpha\,g_A^2}{96\,f_\pi^2\,m_\pi\,\pi}
           +\delta_\alpha\,,\nonumber\\
\beta_M^s &=\frac{\alpha\,g_A^2}{192\,f_\pi^2\,m_\pi\,\pi}
           +\delta_\beta\,.
\label{deltaalpha}
\end{align}
The energy dependence of the polarizabilities is still given solely 
by the leading-order pion cloud.
Eq.~(\ref{deltaalpha}) promotes the short-distance contribution to $\alpha$ 
and $\beta$ from $\mathcal{O}(q^4)$ to $\mathcal{O}(q^3)$. There are 
indications that this change in the power counting is necessary if high-energy
modes in the pion-loop graphs that generate $\alpha$ and $\beta$ are to be 
properly accounted for~\cite{Holstein}.
In order to avoid confusion we denote the fits done with this procedure
as HB$\chi$PT $\mathcal{O}(\bar{q}^3)$.

Fits similar to our $\mathcal{O}(\epsilon^3)$ and $\mathcal{O}(\bar{q}^3)$ 
ones have already been performed in~\cite{McGPhil}, calculating in HB$\chi$PT 
up to $\mathcal{O}(q^4)$. The authors of~\cite{McGPhil} used all available 
data sets but had to exclude the two 94.2~MeV data points measured in the 
backward direction.
As \cite{Phillips,McGPhil} and the $\mathcal{O}(\epsilon^3)$ SSE calculation 
obviously have problems to describe the normalization of the data at
energies below 60~MeV 
we decided to only include the data around 
69~MeV~\cite{Lucas,Lund} and 94.2~MeV~\cite{Hornidge} in the fit. 
We do not make any cuts on the angles and,  
in contradistinction to~\cite{McGPhil}, 
we do not allow the normalizations in the various experiments to float 
in the fit within their quoted systematic errors.

We performed the fits using the NNLO chiral wave function. We fitted the 16 
data points using 2 free parameters ($\alpha_E^s$ and $\beta_M^s$), leaving us 
with 14 degrees of freedom. The resulting values for $\alpha_E^s$ and 
$\beta_M^s$ (see Table~\ref{tablefulldata})  are
\beq
\alpha_E^s=(12.1\pm1.3)\cdot10^{-4}\;\mathrm{fm}^3,\;\;
\beta_M^s =( 1.8\pm1.6)\cdot10^{-4}\;\mathrm{fm}^3
\label{bestresults}
\eeq
with a $\chi^2/d.o.f.$ of 1.78. The corresponding plots are displayed in 
Fig.~\ref{errorfitsfull}, together with the results of our 
$\mathcal{O}(\bar{q}^3)$ fits. 
Using the experimental values from Eq.~(\ref{expp})~\cite{Olmos} as input 
or, equivalently, the values given in Eq.~(\ref{exppHGHP}), which are obtained 
from proton Compton scattering data within the same framework as we are 
using here, 
one can derive the neutron polarizabilities from the isoscalar ones: 
\beq
\alpha_E^n=(12.1\pm1.3)\cdot10^{-4}\;\mathrm{fm}^3,\;\;
\beta_M^n =( 2.0\pm1.6)\cdot10^{-4}\;\mathrm{fm}^3
\eeq
From these results we deduce that the 
isovector polarizabilities are 
rather small (see Table~\ref{tablefulldata}), in good agreement with $\chi$PT
expectations, which predict the 
isovector part to be of higher than third order. Therefore we find no 
contradiction between the results from quasi-free~\cite{Kossert} and elastic 
deuteron Compton scattering.

Our results for $\alpha_E^s$ and $\beta_M^s$ in SSE 
(cf. Eq.(\ref{bestresults})) 
are well consistent (within error bars) with the isoscalar Baldin Sum Rule 
\beq
\alpha_E^s+\beta_M^s\bigg|_{\mathrm{world}\;\, \mathrm{av.}}=
 (14.5\pm0.6)\cdot10^{-4}\;\mathrm{fm}^3,
\label{Baldin}
\eeq
which has been a serious problem in former 
extractions~\cite{Lvov,McGPhil}. The numerical value for the sum rule 
is derived from 
\begin{align}
\alpha_E^p+\beta_M^p&=(13.8\pm0.4)\cdot10^{-4}\;\mathrm{fm}^3\;\;\;\;\;
 \cite{Olmos},\nonumber\\
\alpha_E^n+\beta_M^n&=(15.2\pm0.5)\cdot10^{-4}\;\mathrm{fm}^3\;\;\;\;\;
 \cite{Kossert}.
\end{align}
Due to the consistency of our fit results with the sum rule value from 
Eq.~(\ref{Baldin}) one can in a second step use this number -- we use the 
central value -- as an additional fit constraint and thus reduce the number 
of free parameters to one. The resulting one-parameter fits in SSE of 
Table~\ref{tablefulldata},
\beq
\alpha_E^s=(12.3\pm0.7)\cdot10^{-4}\;\mathrm{fm}^3,\;\;
\beta_M^s =( 2.2\mp0.7)\cdot10^{-4}\;\mathrm{fm}^3,
\eeq
are in good agreement with the isoscalar average of the numbers from 
Eqs.~(\ref{expp}) and~(\ref{expn}) -- or, alternatively, Eqs.~(\ref{expp})
and~(\ref{expSchmied}).

Comparing our fit results to 
the isoscalar $\mathcal{O}(q^4)$ HB$\chi$PT estimate~\cite{BKMS},
 $\alpha_E^s=(11.95\pm2.5)\cdot10^{-4}\;\mathrm{fm}^3$,
 $\beta_M^s = (5.65\pm5.1)\cdot10^{-4}\;\mathrm{fm}^3$, 
we see only minor deviations from their value for $\alpha_E^s$. 
However, our values for $\beta_M^s$ are significantly smaller, but no 
meaningful conclusion can be drawn due to the large error bars in the 
$\mathcal{O}(q^4)$ estimate. The reason for the huge error bars in 
the  $\mathcal{O}(q^4)$ HB$\chi$PT numbers of Ref.~\cite{BKMS} is their 
sensitivity to short-distance contributions which were estimated using the 
resonance-saturation hypothesis.

\subsection{Wave-Function Dependence of the Fits}
\label{fits:wfdep}

To have an estimate on the systematic error due to the wave-function 
dependence, we show our results when we use the two extreme wave 
functions (cf. Fig.~\ref{wavefdep}) for the fit: the NNLO 
chiral wave function~\cite{Epelbaum} and the wave function from the Nijm93 
potential~\cite{Nijm}. Furthermore, we are fitting in two different ways:
First the number of degrees of freedom is the number of data points (16) minus 
the number of free parameters (2).  In a second step we use the isoscalar 
Baldin Sum Rule, Eq.~(\ref{Baldin}), to reduce the number of degrees of 
freedom to 1, as described before.

Fitting the $\gamma d$ cross sections with the $\mathcal{O}(\epsilon^3)$ and 
$\mathcal{O}(\bar{q}^3)$ kernel, respectively, using the Nijm93 wave 
function yields larger results for $\alpha_E$ and smaller ones for 
$\beta_M$, but still the values of Table~\ref{tablefulldata} are in 
reasonable agreement with the values given 
in Eq.~(\ref{expn})~\cite{Kossert}. Comparing the 
differing results that we get for $\alpha_E^s$ with the Nijm93 and the  
NNLO $\chi$PT wave function,
we estimate our systematic error to be of the order of 15$\,$\%. 

\begin{table}[!htb]
\begin{center}
\begin{tabular}{|c||c||c|c|c|c|}
\hline
Amplitudes&Quantity&2-par. fit    &1-par. fit   &2-par. fit&1-par. fit\\
          &        &NNLO $\chi$PT &NNLO $\chi$PT&Nijm93    &Nijm93    \\    
\hline
\hline
$\mathcal{O}(\epsilon^3)$~SSE&$\chi^2/d.o.f.$ &1.78&1.67&2.45&2.35\\
\cline{2-6}
&$\alpha_E^s\;\,[10^{-4}\,\mathrm{fm}^3]$&$12.1\pm1.3$
                 &$12.3\pm0.7$&$14.3\pm 1.3$&$13.8\pm0.7$\\
&$\beta _M^s\;\,[10^{-4}\,\mathrm{fm}^3]$&$ 1.8\pm1.6$
                 &$ 2.2\mp0.7$&$ 1.5\pm 1.6$&$ 0.7\mp0.7$\\
\cline{2-6}
&$\alpha_E^s+\beta _M^s\;\,[10^{-4}\,\mathrm{fm}^3]$
                       &$13.9\pm2.1$&$14.5\,(\mathrm{fit})$
                       &$15.8\pm2.1$&$14.5\,(\mathrm{fit})$\\
\cline{2-6}
&$\alpha_E^n\;\,[10^{-4}\,\mathrm{fm}^3]$&$12.1\pm1.3$
                 &$12.5\pm0.8$&$16.5\pm 1.3$&$15.5\pm0.8$\\
&$\beta _M^n\;\,[10^{-4}\,\mathrm{fm}^3]$&$ 2.0\pm1.6$
                 &$ 2.8\pm0.8$&$ 1.4\pm 1.6$&$ -0.2\pm0.8$\\
\hline
\hline
$\mathcal{O}(\bar{q}^3)$~HB$\chi$PT&$\chi^2/d.o.f.$ &2.14&2.01&2.87&2.75\\
\cline{2-6}
&$\alpha_E^s\;\,[10^{-4}\,\mathrm{fm}^3]$&$11.0\pm1.3$
                 &$11.3\pm0.7$&$13.2\pm 1.2$&$12.7\pm0.7$\\
&$\beta _M^s\;\,[10^{-4}\,\mathrm{fm}^3]$&$ 2.8\pm1.6$
                 &$ 3.2\mp0.7$&$ 2.5\pm 1.5$&$ 1.8\mp0.7$\\
\cline{2-6}
&$\alpha_E^s+\beta _M^s\;\,[10^{-4}\,\mathrm{fm}^3]$
                       &$13.8\pm2.1$&$14.5\,(\mathrm{fit})$
                       &$15.7\pm1.9$&$14.5\,(\mathrm{fit})$\\
\cline{2-6}
&$\alpha_E^n\;\,[10^{-4}\,\mathrm{fm}^3]$&$9.9\pm1.3$
                 &$10.5\pm0.8$&$14.3\pm 1.2$&$13.3\pm0.8$\\
&$\beta _M^n\;\,[10^{-4}\,\mathrm{fm}^3]$&$ 4.0\pm1.6$
                 &$ 4.8\pm0.8$&$ 3.4\pm 1.5$&$ 2.0\pm0.8$\\
\hline
\end{tabular}
\caption{Values for 
the isoscalar and neutron polarizabilities from a fit to the full 69~MeV and 
94.2~MeV data sets~\cite{Lucas, Lund, Hornidge}, using the 
$\mathcal{O}(\epsilon^3)$~SSE amplitudes and the 
$\mathcal{O}(\bar{q}^3)$~HB$\chi$PT amplitudes, respectively.
The neutron results are derived 
using the proton values from~\cite{Olmos} (Eq.~(\ref{expp})) as input.
All error bars displayed are only statistical.}
\label{tablefulldata}
\end{center}
\end{table}

One of the reasons for the differing 
results between our approach and the calculations presented 
in~\cite{Lvov,Miller}  
is the energy dependence of the polarizabilities. 
In our calculation, it is completely given by the Compton multipoles, whereas 
the authors of~\cite{Lvov,Miller} only used the leading~\cite{Lvov,Miller} and 
subleading~\cite{Lvov} terms of a Taylor expansion
of $\alpha_E(\omega)$ and $\beta_M(\omega)$ in the photon energy $\omega$. 
Comparing the range of our results for $\alpha_E^s$ and $\beta_M^s$ to the 
ranges quoted in~\cite{McGPhil} (cf. Sect.~\ref{sec:introduction}), 
we observe a slight tendency towards larger 
values of $\alpha_E^s$ and smaller ones of $\beta_M^s$ in our analysis. 
The reason for this deviation will be discussed in detail in 
Sect.~\ref{sec:unbiased}.

\subsection{Comparison of $\mathcal{O}(\epsilon^3)$ SSE and 
                          $\mathcal{O}(\bar{q}^3)$ HB$\chi$PT Fits}
\label{sec:Comparison}

When we compare the $\mathcal{O}(\epsilon^3)$ SSE fit results for 
$\alpha_E^s$ and $\beta_M^s$ with the corresponding 
$\mathcal{O}(\bar{q}^3)$ HB$\chi$PT results (Table~\ref{tablefulldata}), 
we see that in the HB$\chi$PT fit (Eq.~(\ref{deltaalpha})) the electric 
dipole polarizability is smaller, whereas $\beta_M^s$ turns out to be larger. 
The reason for the systematic shift of the magnetic polarizability is that 
due to the missing $\Delta(1232)$ resonance in HB$\chi$PT the static value of 
$\beta_M$ is inflated in order to compensate for the paramagnetic rise of the 
resonance. This will be discussed further in the next subsection.

\begin{figure}[!htb]
\begin{center} 
\includegraphics*[width=.48\textwidth]
{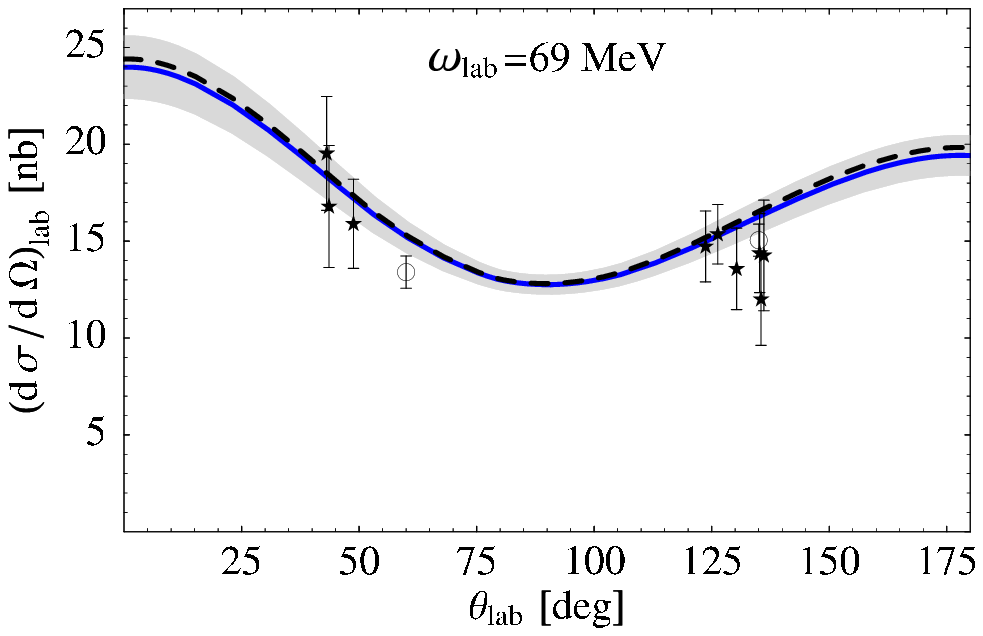}
\hfill
\includegraphics*[width=.48\textwidth]
{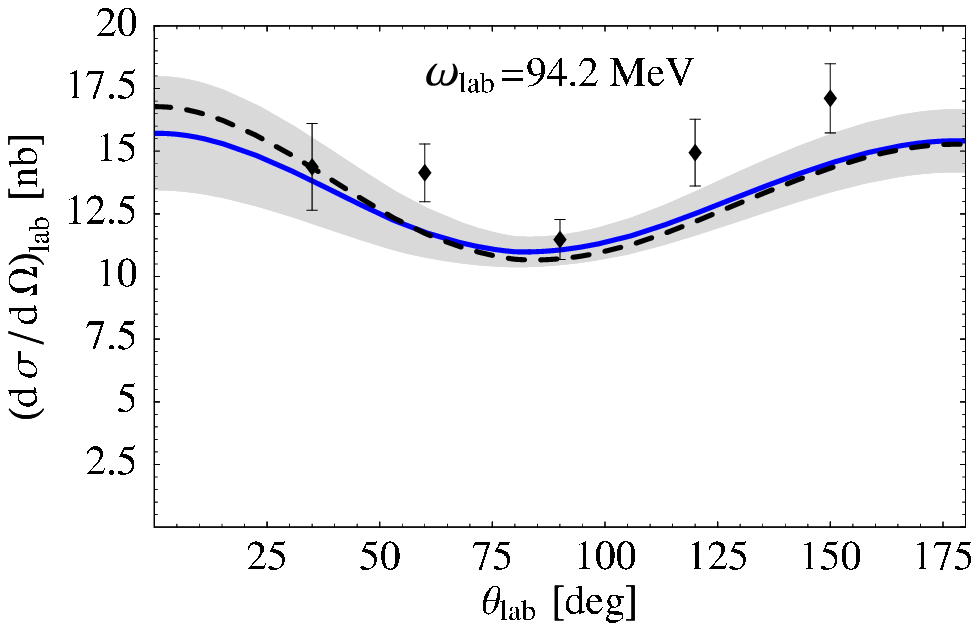}
\includegraphics*[width=.48\textwidth]
{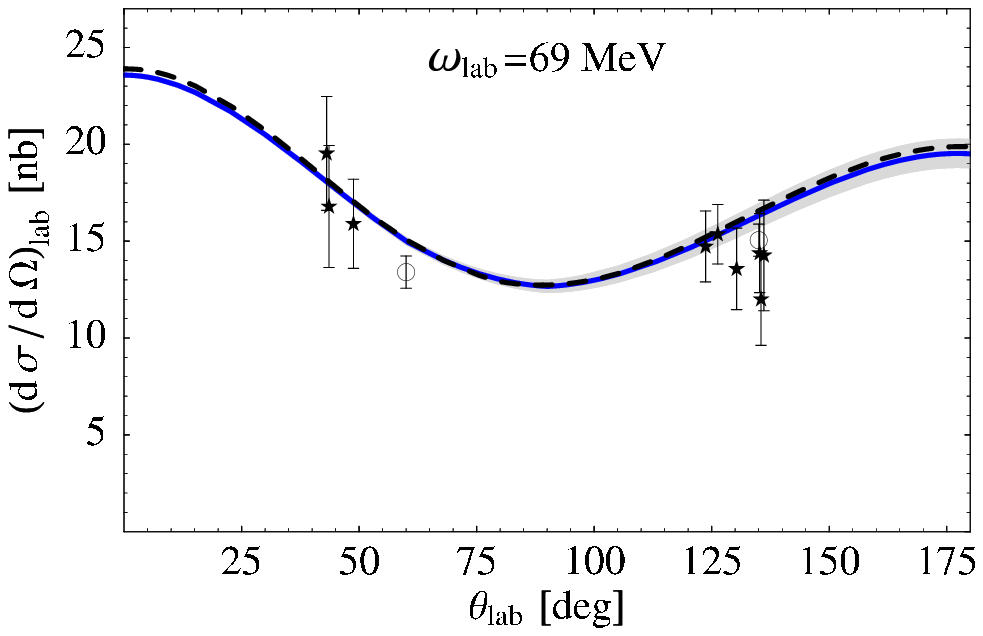}
\hfill
\includegraphics*[width=.48\textwidth]
{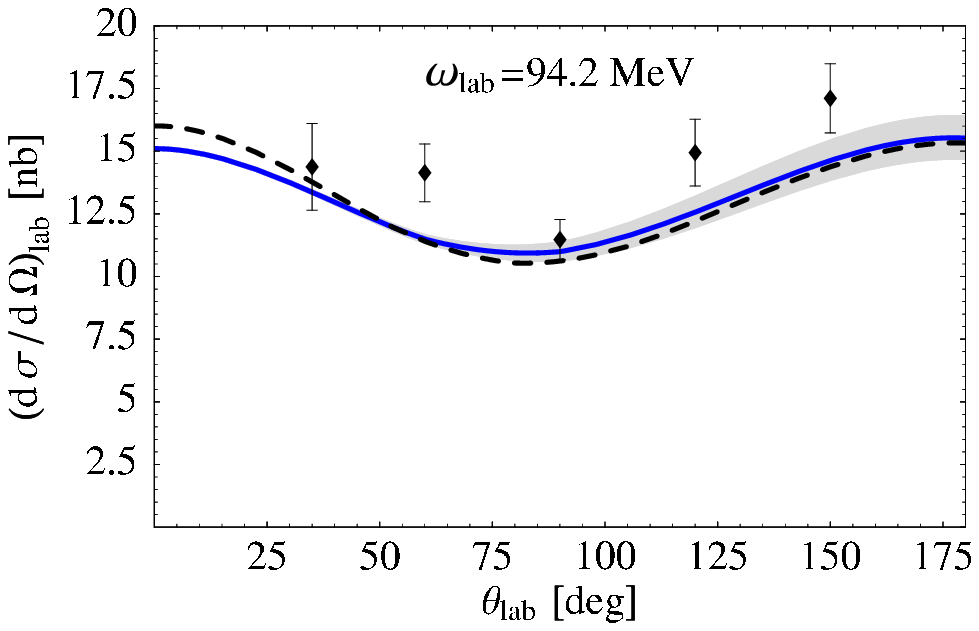}
\parbox{1.\textwidth}{
\caption{$\mathcal{O}(\epsilon^3)$ SSE (solid) and 
$\mathcal{O}(\bar{q}^3)$ HB$\chi$PT (dashed) results with $\alpha_E^s$, 
$\beta_M^s$ from Table~\ref{tablefulldata}, using
the chiral NNLO wave function~\cite{Epelbaum}. 
The upper panels correspond 
to a fit of both polarizabilities, in the lower panels the Baldin Sum Rule 
(cf. Eq.~(\ref{Baldin})) is 
used as additional fit constraint. The grey bands are derived from our 
(statistical) errors.}
\label{errorfitsfull}}
\end{center}
\end{figure}
Fig.~\ref{errorfitsfull} demonstrates that this compensation works very well 
in the $\gamma d$ cross sections, as 
the curves, which correspond to the $\mathcal{O}(\epsilon^3)$ SSE and to the 
$\mathcal{O}(\bar{q}^3)$ HB$\chi$PT fits, are nearly indistinguishable.
We consider the quality of our $\mathcal{O}(\bar{q}^3)$ fit to be 
comparable to that of the $\mathcal{O}(q^4)$ fits of Ref.~\cite{McGPhil}. 

The SSE and HB$\chi$PT fit results shown in Fig.~\ref{errorfitsfull} only 
differ in the associated pairs $\alpha_E$,~$\beta_M$. 
Therefore, from the available $\gamma d$ 
data alone one cannot draw any firm conclusion regarding the 
importance of explicit $\Delta(1232)$ degrees of freedom.
However, as we will show in the next section,
from $\gamma p$ scattering experiments it is clear
that third-order HB$\chi$PT does not describe 
the dynamics in the $\gamma d$ process correctly. Given that the SSE 
calculation describes both the $\gamma p$ and the $\gamma d$ experiments we 
believe we have established that a Chiral Effective Field Theory which 
includes the explicit $\Delta$ field is 
an efficient framework with which to describe low-energy Compton scattering.
\subsection{Towards an Unbiased Fitting Procedure}
\label{sec:unbiased}

\begin{table}[!htb]
\begin{center}
\begin{tabular}{|c||c|}
\hline
angle [deg]&$\mathrm{d}\sigma/\mathrm{d}\Omega$ [nb]\\
\hline
 45.6&$17.3\pm2.8$\\
130.5&$14.2\pm2.1$\\
\hline
\end{tabular}
\caption{Effective data points representing the 67~MeV data from~\cite{Lund}.}
\label{tab:eff}
\end{center}
\end{table}

\begin{figure}[!htb]
\begin{center} 
\includegraphics*[width=.48\textwidth]
{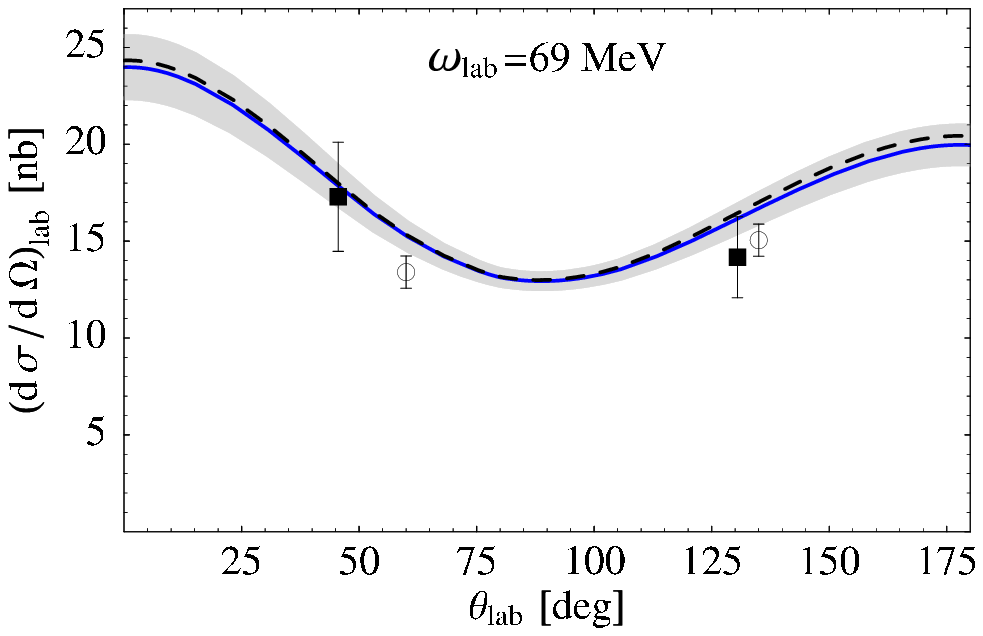}
\hfill
\includegraphics*[width=.48\textwidth]
{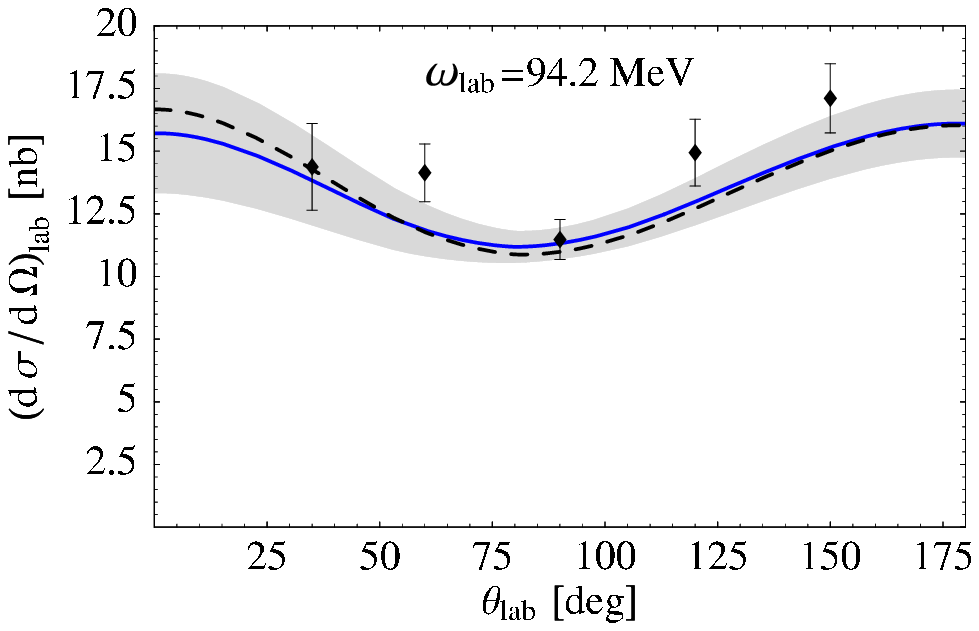}
\includegraphics*[width=.48\textwidth]
{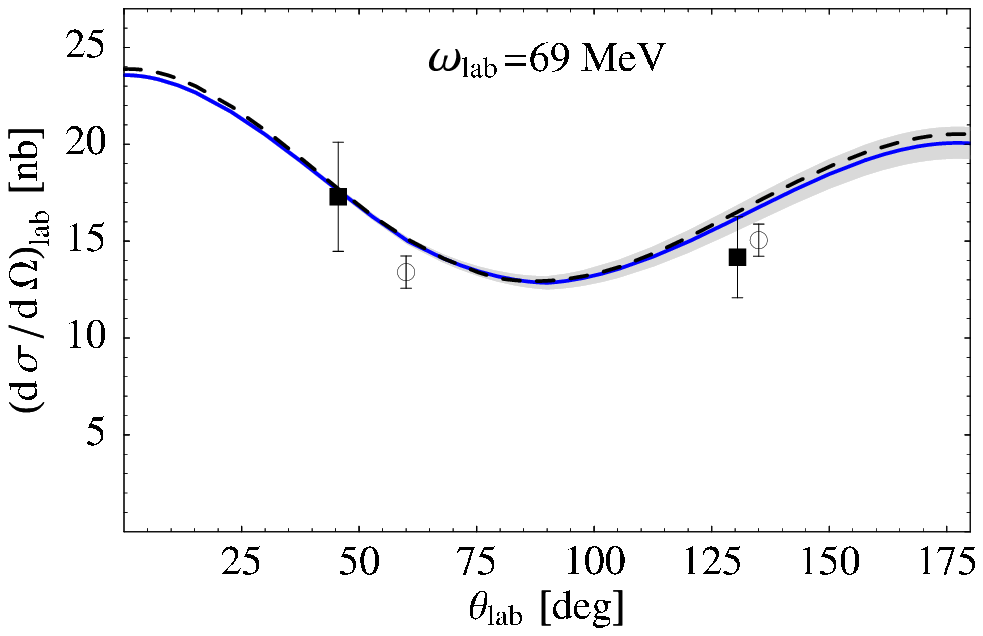}
\hfill
\includegraphics*[width=.48\textwidth]
{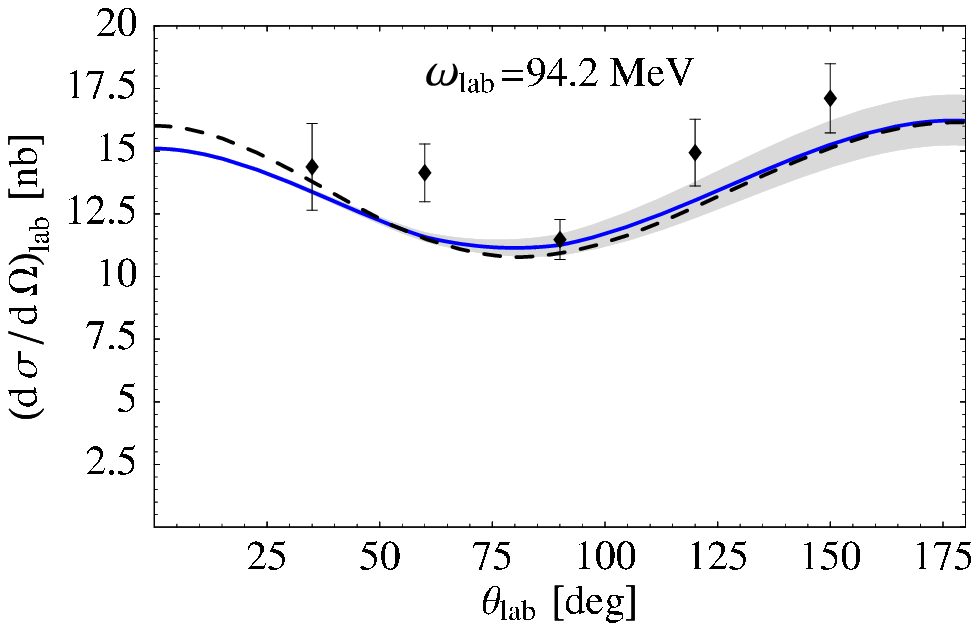}
\parbox{1.\textwidth}{
\caption{$\mathcal{O}(\epsilon^3)$ SSE (solid) and 
$\mathcal{O}(\bar{q}^3)$ HB$\chi$PT (dashed) results with $\alpha_E^s$, 
$\beta_M^s$ from Table~\ref{tableeffdata}, using
the chiral NNLO wave function~\cite{Epelbaum}. 
The upper panels correspond 
to a fit of both polarizabilities, in the lower panels the Baldin Sum Rule 
(cf. Eq.~(\ref{Baldin})) is 
used as additional fit constraint. The grey bands are derived from our 
(statistical) errors. The two 
data points plotted as boxes are the effective data that we use as 
representative pseudo-data points for the data from~\cite{Lund}. The full 
set of data points can be seen in Fig.~\ref{errorfitsfull}.}
\label{errorfitseff}}
\end{center}
\end{figure}

When we compare the several data sets that we use for the fits, we find 
eleven data points at $\omega_\mathrm{lab}\approx 69$~MeV, centered around 
only two different angles, and five points at 
$\omega_\mathrm{lab}\approx 94.2$~MeV, distributed 
over the whole angular spectrum. Especially around 
$\theta_\mathrm{lab} \approx 130^\circ$ there is a wealth of data around 
69~MeV (six points from \cite{Lund} and one from \cite{Lucas}), which gives an 
anomalously strong constraint to our fit routines. As long as there are no 
further data available at higher energies, fitting to all of the 69~MeV data 
thus overestimates this 
energy region with respect to the 94.2~MeV data from \cite{Hornidge}.
Therefore, in the following we compensate for this imbalance of data by 
replacing the Lund data~\cite{Lund} by two ``effective'' pseudo-data points 
(cf. Table~\ref{tab:eff} and
Fig.~\ref{errorfitseff}), which represent the data in the forward and backward 
direction, respectively. These are obtained by weighting the angles and the 
differential cross section values of the represented data points by the inverse
 of their errors, and we assign the average over the errors of the 
represented data as error bars. Therefore, the remaining data are the two data 
points from~\cite{Lucas} at 69~MeV, the two ``effective'' data at 
$\sim$67~MeV, shown in Table~\ref{tab:eff}, 
representing~\cite{Lund}, and the five data points 
from~\cite{Hornidge} around 94.2~MeV. With these 
nine data points we perform the same fits as we did before for the complete 
data sets.
The resulting values for $\alpha_E$ and $\beta_M$ are presented in 
Table~\ref{tableeffdata}, the plots (including the two effective data points)
in Fig.~\ref{errorfitseff}, exhibiting better agreement with 
the 94.2~MeV data than the fits of Sect.~\ref{sec:Comparison} 
(Fig.~\ref{errorfitsfull}), as expected.
Comparing the results for $\alpha_E^s$ and 
$\beta_M^s$ of Table~\ref{tableeffdata} (or, equivalently, 
for $\alpha_E^n$ and $\beta_M^n$) to the 
results  from our fits to all data points, given in Table~\ref{tablefulldata}, 
we note that both for $\mathcal{O}(\epsilon^3)$ SSE and 
$\mathcal{O}(\bar{q}^3)$ HB$\chi$PT 
$\alpha_E$ is slightly smaller ($\beta_M$ slightly larger). 

\begin{figure}[!htb]
\begin{center} 
\includegraphics*[width=.48\textwidth]{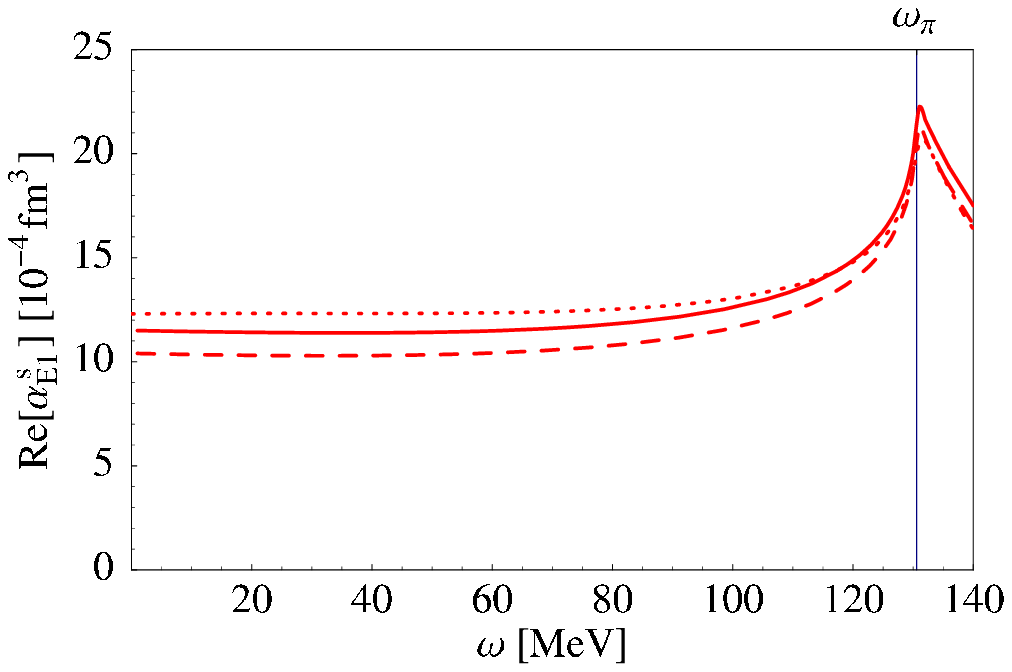}
\hfill
\includegraphics*[width=.48\textwidth]{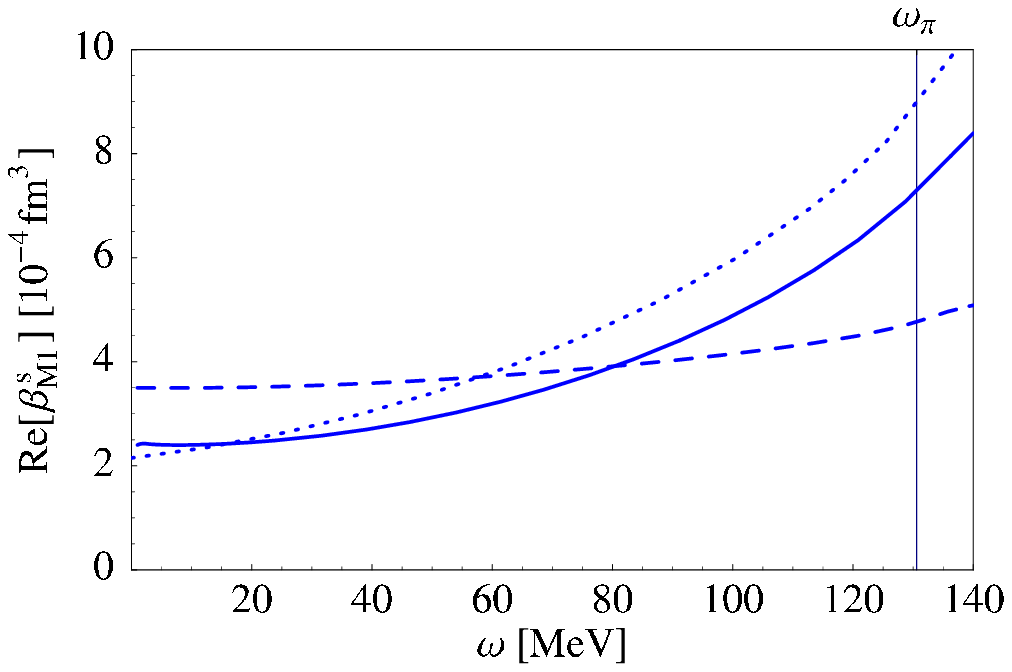}
\parbox{1.\textwidth}{
\caption{Comparison of the isoscalar Dispersion Theory result (dotted) of 
Ref.~\cite{HGHP} for the dynamical dipole polarizabilities 
$\alpha_{E1}^s(\omega)$ and $\beta_{M1}^s(\omega)$ to the 
$\mathcal{O}(\epsilon^3)$ SSE (solid) and the
$\mathcal{O}(\bar{q}^3)$ HB$\chi$PT (dashed) results with the static 
values from the two-parameter fit using the chiral wave function, given 
in Table~\ref{tableeffdata}. 
$\omega_\pi$ denotes the pion-production threshold.}
\label{dynpolas}}
\end{center}
\end{figure}

We also see once again that the theory 
without explicit $\Delta$ degrees of freedom leads to a 
systematically larger value for $\beta_M^s$, supporting our hypothesis that 
the enhancement is due to the insufficient dynamics in the HB$\chi$PT (cf. 
Table~\ref{tablefulldata}).
This is demonstrated in Fig.~\ref{dynpolas}, which shows the dynamical 
isoscalar dipole polarizabilities 
$\alpha_{E1}^s(\omega)$, $\beta_{M1}^s(\omega)$, calculated from 
third-order SSE and HB$\chi$PT, respectively. The static values are taken from
 the unconstrained fit, using the NNLO $\chi$PT wave function, i.e. 
\begin{equation}
\alpha_{E}^s=11.5\cdot10^{-4}\;\mathrm{fm}^3\,,\;\;\;\; 
\beta_{M}^s = 2.4\cdot10^{-4}\;\mathrm{fm}^3 
\end{equation}
for the $\mathcal{O}(\epsilon^3)$ SSE-curve, and 
\begin{equation}
\alpha_{E}^s=10.4\cdot10^{-4}\;\mathrm{fm}^3\,,\;\;\;\; 
\beta_{M}^s = 3.5\cdot10^{-4}\;\mathrm{fm}^3
\end{equation}
for the $\mathcal{O}(\bar{q}^3)$ $\chi$PT-curve. 
As shown in Fig.~\ref{dynpolas} the energy dependence of the two 
field-theoretical calculations 
for $\alpha_{E1}(\omega)$ is in good agreement with each other and with the 
recent analysis from Dispersion Theory~\cite{HGHP}.  
Matters are different for $\beta_{M1}(\omega)$: Whereas the SSE-curve 
reproduces the paramagnetic rise 
due to the explicitly included $\Delta$ resonance,
the HB$\chi$PT result amounts to a nearly energy-independent average for this 
quantity. However, it is well-known that this paramagnetic rise in the 
Compton multipoles is necessary for the correct description of the $\gamma p$ 
data around pion threshold, as can also be seen in the Dispersion-Theory 
analysis. In HB$\chi$PT the 
static value is artificially enhanced by the fit constraint from the 94.2~MeV 
data in order to compensate for the missing dynamics. 
Both in Fig.~\ref{errorfitsfull} and Fig.~\ref{errorfitseff} the enhanced 
$\beta_M$ is able to cure the $\gamma d$ cross 
sections and make the resulting curves very similar to the plots from the 
SSE fits (cf. Fig.~\ref{errorfitseff}). 

Therefore, for understanding the 
available $\gamma d$ data via fits of $\alpha_E^s$ and $\beta_M^s$, 
it is essential to combine the pairs $\alpha_E$,~$\beta_M$ resulting from the 
$\gamma d$ analysis with an energy-dependent multipole analysis of $\gamma p$ 
scattering. 
From the information available on Compton multipoles from $\gamma p$ 
scattering experiments, it is clear that third-order HB$\chi$PT is too 
simplistic a picture for the dynamics of the $\gamma d$ process at energies of 
$\mathcal{O}(100\;\mathrm{MeV})$. It is therefore
crucial that deuteron and proton Compton experiments are available at 
comparable energies and that they are analyzed within the same framework.

Putting equal statistical weight on the 69 and the 94.2~MeV data
can be seen as a demonstration of the importance of obtaining 
comparable statistics at all energies.
We therefore urge for more experimental information at photon energies 
around 100~MeV.
With such information, deuteron Compton cross sections below the pion mass 
provide an excellent window to investigate which internal nucleonic degrees 
of freedom contribute in both processes, $\gamma p\rightarrow \gamma p$ and 
$\gamma d\rightarrow \gamma d$.  

\begin{table}[!htb]
\begin{center}
\begin{tabular}{|c||c||c|c|c|c|}
\hline
Amplitudes&Quantity&2-par. fit    &1-par. fit   &2-par. fit&1-par. fit\\
          &        &NNLO $\chi$PT &NNLO $\chi$PT&Nijm93    &Nijm93    \\    
\hline
\hline
$\mathcal{O}(\epsilon^3)$~SSE&$\chi^2/d.o.f.$ &2.61&2.28&3.72&3.37\\
\cline{2-6}
&$\alpha_E^s\;\,[10^{-4}\,\mathrm{fm}^3]$&$11.5\pm1.4$
                 &$11.7\pm0.8$&$13.6\pm 1.4$&$13.1\pm0.7$\\
&$\beta _M^s\;\,[10^{-4}\,\mathrm{fm}^3]$&$ 2.4\pm1.7$
                 &$ 2.8\mp0.8$&$ 2.2\pm 1.7$&$ 1.4\mp0.7$\\
\cline{2-6}
&$\alpha_E^s+\beta _M^s\;\,[10^{-4}\,\mathrm{fm}^3]$ 
                       &$13.9\pm2.2$&$14.5\,(\mathrm{fit})$
                       &$15.8\pm2.2$&$14.5\,(\mathrm{fit})$\\
\cline{2-6}
&$\alpha_E^n\;\,[10^{-4}\,\mathrm{fm}^3]$&$10.9\pm1.4$
                 &$11.3\pm0.9$&$15.1\pm 1.4$&$14.1\pm0.8$\\
&$\beta _M^n\;\,[10^{-4}\,\mathrm{fm}^3]$&$ 3.2\pm1.7$
                 &$ 4.0\pm0.9$&$ 2.8\pm 1.7$&$ 1.2\pm0.8$\\
\hline
\hline
$\mathcal{O}(\bar{q}^3)$~HB$\chi$PT&$\chi^2/d.o.f.$ &3.14&2.77&4.36&3.93\\
\cline{2-6}
&$\alpha_E^s\;\,[10^{-4}\,\mathrm{fm}^3]$&$10.4\pm1.3$
                 &$10.6\pm0.8$&$12.4\pm 1.3$&$12.0\pm0.8$\\
&$\beta _M^s\;\,[10^{-4}\,\mathrm{fm}^3]$&$ 3.5\pm1.7$
                 &$ 3.9\mp0.8$&$ 3.3\pm 1.6$&$ 2.5\mp0.8$\\
\cline{2-6}
&$\alpha_E^s+\beta _M^s\;\,[10^{-4}\,\mathrm{fm}^3]$
                       &$13.9\pm2.1$&$14.5\,(\mathrm{fit})$
                       &$15.7\pm2.1$&$14.5\,(\mathrm{fit})$\\
\cline{2-6}
&$\alpha_E^n\;\,[10^{-4}\,\mathrm{fm}^3]$&$8.7\pm1.3$
                 &$9.1\pm0.9$&$12.7\pm 1.3$&$11.9\pm0.9$\\
&$\beta _M^n\;\,[10^{-4}\,\mathrm{fm}^3]$&$ 5.4\pm1.7$
                 &$ 6.2\pm0.9$&$ 5.0\pm 1.6$&$ 3.4\pm0.9$\\
\hline
\end{tabular}
\caption{Values for 
the isoscalar and neutron polarizabilities from a fit to the 69~MeV and 
94.2~MeV data sets~\cite{Lucas, Hornidge}, using the 
$\mathcal{O}(\epsilon^3)$~SSE amplitudes and the 
$\mathcal{O}(\bar{q}^3)$~HB$\chi$PT
amplitudes, respectively. The data from~\cite{Lund} have been replaced by two 
pseudo-data points, specified in Table~\ref{tab:eff}.
The neutron results are derived 
using the proton values from~\cite{Olmos} as input.
All error bars displayed are only statistical.}
\label{tableeffdata}
\end{center}
\end{table}

\section{Conclusion and Outlook} 
\label{sec:outlook}

In this work Compton scattering from the deuteron was calculated 
up to next-to-leading order in the Small Scale Expansion, an Effective
Field Theory with nucleons, pions and the $\Delta(1232)$ resonance as explicit
degrees of freedom. We investigated three different energies and compared to 
the available experimental data, finding good agreement for 69~MeV and 94.2~MeV
and a failure of our calculation for 49~MeV. 
The reason for this last result is that for low energies our power counting 
breaks down, so one 
has to modify the power counting in the very-low-energy region and use 
non-perturbative methods in order to reproduce the correct low-energy 
theorems. 
It will be one of our future 
projects~\cite{future} to get the region $\omega<50$~MeV under control 
and finally to restore the correct Thomson limit. 
Here we concentrated on the energy range between 50~MeV and 100~MeV. We 
found that our calculation gives reasonable results for photon energies above 
60~MeV. Motivated by the good agreement at these higher energies, we fitted the
isoscalar polarizabilities $\alpha_E^s$ and $\beta_M^s$ to the data 
around 69~MeV and 94.2~MeV, yielding results in good agreement with the 
$\mathcal{O}(q^4)$ HB$\chi$PT estimate of Ref.~\cite{BKMS} and experiment. 
Averaging over the results of our two
unconstrained SSE fits (one with the chiral NNLO wave function~\cite{Epelbaum},
 one with the Nijm93 wave function~\cite{Nijm}, cf. Table~\ref{tablefulldata}) 
results in the isoscalar polarizabilities
\begin{align}
\alpha_E^s&=(13.2\pm1.3\,(\mathrm{stat})\pm2.1\,(\mathrm{syst}))
            \cdot 10^{-4}\;\mathrm{fm}^3\,,\nonumber\\
\beta_M^s &=( 1.7\pm1.6\,(\mathrm{stat})\pm0.3\,(\mathrm{syst}))
            \cdot 10^{-4}\;\mathrm{fm}^3\,,
\label{eq:conclusion}
\end{align}
where we assumed the same statistical errors as in Table~\ref{tablefulldata}. 
The systematic error 
due to the differing results when we use different wave functions was 
estimated to be around 15$\,$\%. 
As these results are in good agreement with the isoscalar Baldin Sum Rule, 
cf. Eq.~(\ref{Baldin}), we also used the central sum-rule 
value as additional fit constraint, obtaining
\begin{align}
\alpha_E^s&=(13.1\pm0.7\,(\mathrm{stat})\pm2.0\,(\mathrm{syst}))
            \cdot 10^{-4}\;\mathrm{fm}^3\,,\nonumber\\
\beta_M^s &=( 1.5\mp0.7\,(\mathrm{stat})\pm0.2\,(\mathrm{syst}))
            \cdot 10^{-4}\;\mathrm{fm}^3\,.
\label{eq:conclusionBaldin}
\end{align}

Motivated by the statistical imbalance between experimental data around 
94.2~MeV and 69~MeV, we reduced in a second fit the statistics at 69~MeV, 
replacing the nine data points given in~\cite{Lund} by two representative
pseudo-data points, leading to an equal weighting between the two energy sets.
The bias-corrected fitting procedure confirms our findings of small values 
for $\beta_M^s$, implying small isovector components. 
Our results for the isoscalar polarizabilities that we derive from the fit 
including only the two representatives of the data from~\cite{Lund} are
\begin{align}
\alpha_E^s&=(12.6\pm1.4\,(\mathrm{stat})\pm1.9\,(\mathrm{syst}))
            \cdot 10^{-4}\;\mathrm{fm}^3\,,\nonumber\\
\beta_M^s &=( 2.3\pm1.7\,(\mathrm{stat})\pm0.3\,(\mathrm{syst}))
            \cdot 10^{-4}\;\mathrm{fm}^3\,.
\label{eq:conclusionunbiased}
\end{align}
Including the Baldin constraint we get
\begin{align}
\alpha_E^s&=(12.4\pm0.8\,(\mathrm{stat})\pm1.9\,(\mathrm{syst}))
            \cdot 10^{-4}\;\mathrm{fm}^3\,,\nonumber\\
\beta_M^s &=( 2.1\mp0.8\,(\mathrm{stat})\pm0.3\,(\mathrm{syst}))
            \cdot 10^{-4}\;\mathrm{fm}^3\,.
\label{eq:conclusionunbiasedBaldin}
\end{align}
We consider these ``unbiased'' results, 
Eqs.~(\ref{eq:conclusionunbiased}) and~(\ref{eq:conclusionunbiasedBaldin}), 
to be the more reliable values, 
since in a straightforward fit to the existing $\gamma d$ data
there is an obvious imbalance between the number of points at 69 and 94.2~MeV. 
The data base should be enlarged at higher energies so that  
unbiased fit results can be obtained. If further experiments, as planned 
at TUNL/HI$\gamma$S or at MAXlab, provide additional data at energies 
between 70~MeV and the pion mass over the whole angular range, an unbiased 
fitting routine including 
data over this entire energy region will be possible. However, we caution 
that we found a very strong 
energy dependence of the $\gamma d$ cross sections in the forward direction.
Therefore, we would recommend that any future data taken over a range of 
photon energies be analyzed using a model which incorporates this rapid 
energy dependence.

A previous analysis of $\gamma p$ data within the same chiral Effective Field 
Theory used here yielded the values 
$\alpha_E^p=(11.04\pm1.36)\cdot 10^{-4}\;\mathrm{fm}^3$,
$\beta_M^p =( 2.76\mp1.36)\cdot 10^{-4}\;\mathrm{fm}^3$ 
for the proton polarizabilities~\cite{HGHP}, cf. Eq.~(\ref{exppHGHP}), 
which are consistent with experimental values~\cite{Olmos}.
Combining the numbers of Eq.~(\ref{eq:conclusionunbiased}) with these 
results, we obtain a consistent Effective Field Theory determination 
of the neutron polarizabilities with a precision 
comparable to~\cite{Kossert}:
\begin{align}
\alpha_E^n&=(14.2\pm2.0\,(\mathrm{stat})\pm1.9\,(\mathrm{syst})
            )\cdot 10^{-4}\;\mathrm{fm}^3\,\nonumber\\
\beta_M^n &=( 1.8\pm2.2\,(\mathrm{stat})\pm0.3\,(\mathrm{syst})
            )\cdot 10^{-4}\;\mathrm{fm}^3\,
\label{con:neutron}
\end{align} 
Eq.~(\ref{con:neutron}) does not include the Baldin Sum Rule, whereas the 
one-parameter fit using the Baldin constraint gives
\begin{align}
\alpha_E^n&=(13.8\pm1.6\,(\mathrm{stat})\pm2.1\,(\mathrm{syst})
            )\cdot 10^{-4}\;\mathrm{fm}^3\,,\nonumber\\
\beta_M^n &=( 1.4\mp1.6\,(\mathrm{stat})\pm0.2\,(\mathrm{syst})
            )\cdot 10^{-4}\;\mathrm{fm}^3\,.
\label{con:neutronBaldin}
\end{align} 
It is clear from 
Eqs.~(\ref{exppHGHP}),~(\ref{con:neutron}),~(\ref{con:neutronBaldin}) that the 
isovectorial components -- i.e. the
differences between proton and neutron polarizabilities -- are rather small.
This finding is in good agreement with~\cite{Kossert}, where quasi-elastic 
Compton scattering off the proton and neutron was measured. 
Eqs.~(\ref{con:neutron}) and~(\ref{con:neutronBaldin}) prove that 
small isovectorial nucleon polarizabilities are not in contradiction with 
elastic deuteron Compton scattering data. 
We conclude that both the 
quasi-elastic and the elastic deuteron Compton experiments are consistent 
with small isovectorial polarizabilities.

Furthermore we used the $\mathcal{O}(\bar{q}^3)$ HB$\chi$PT amplitudes for 
analogous fits, finding similar values for $\alpha_{E}$ but larger ones 
for $\beta_{M}$, which is not surprising, as the dynamics of the resonant 
Compton multipoles is not well captured in third-order HB$\chi$PT. Therefore, 
the static value becomes large, since it must correct for the missing 
$\Delta$ resonance, leading
$\mathcal{O}(\bar{q}^3)$ HB$\chi$PT to a disagreement 
with the single-nucleon 
Compton multipoles extracted in theories with an explicit $\Delta(1232)$, e.g. 
Ref.~\cite{HGHP}. Obviously, $\gamma d$ scattering alone is not sufficient
to investigate the relevant low-energy degrees of freedom in nuclear Compton 
scattering, but one has to combine information from $\gamma d$ and $\gamma p$
scattering and analyze both in the same framework.


Finally, in future work one needs to address the issue of the sensitivity to 
the wave function (cf. Sect.~\ref{wfdep}), as one 
would expect that for photon energies below 100~MeV 
any effects of short distance, i.e. high-energy physics, 
should be able to be encoded in counter-terms within a well-understood power 
counting scheme. 


\section*{Acknowledgments}

The authors thank Haiyan Gao and Wolfram Weise for helpful discussions and 
E. Epelbaum and V. Stoks for providing us with their deuteron wave functions.
HWG, TRH and RPH are grateful to the ECT* in Trento for its hospitality. 
HWG thanks the Nuclear Theory Group of Lawrence Berkeley National
Laboratory and the INT in Seattle for their hospitality and
financial support,  instrumental for this research. 
HWG is grateful to the organizers and participants of the 
``Berkeley Visitors Program on Effective Field Theories 2003''. HWG and DRP 
are grateful to the organizers of the 
``INT Program 03-3: Theories of Nuclear Forces and Nuclear Systems'' for the 
financial support.
RPH is grateful to Ohio University, Athens for its hospitality.
This work was supported in part by the Bundesministerium f\"ur
Forschung und Technologie, by the Deutsche Forschungsgemeinschaft under
contracts GR1887/2-2 (HWG, RPH and TRH) and 3-1 (HWG) and by the US DOE 
under grant DE-FG02-02ER41218 and DE-FG02-93ER40756 (DRP). 

\newpage

\appendix

\newpage



\end{document}